\documentclass{aa}%[referee]{aa}
\usepackage[varg]{txfonts}
\usepackage{natbib, amsmath, amssymb, amsfonts, graphicx}
\usepackage{mathtools}
\usepackage{subfig}
\usepackage{comment}
\usepackage{xcolor}
\usepackage[citecolor=blue, linkcolor=blue, urlcolor = black, colorlinks = true]{hyperref}

\title{The traditional approximation of rotation\\ for rapidly rotating stars and planets}
\subtitle{I. The impact of strong deformation}
\author{H. Dhouib \inst{1}
\and V. Prat \inst{1}
\and T. Van Reeth \inst{2}
\and S. Mathis \inst{1}}

\institute{Département d’Astrophysique-AIM, CEA/DRF/IRFU, CNRS/INSU, Université Paris-Saclay, Université Paris-Diderot, Université
de Paris, F-91191 Gif-sur-Yvette, France\\\email{hachem.dhouib@cea.fr}
\and Institute of Astronomy, KU Leuven, Celestijnenlaan 200D, 3001 Leuven, Belgium} 

\titlerunning{The traditional approximation of rotation for rapidly rotating stars and planets. I.}
\authorrunning{H. Dhouib et al.}

\abstract
{The traditional approximation of rotation (TAR) is a treatment of the hydrodynamic equations of rotating and stably stratified fluids in which the action of the Coriolis acceleration along the direction of the entropy and chemical stratifications is neglected because it is weak in comparison with the buoyancy Archimedean force. This leads to the neglect of the horizontal projection of the rotation vector in the equations for the dynamics of gravito-inertial waves (GIWs). The dependent variables in those equations then become separable into radial and horizontal parts as in the non-rotating case. The TAR is built on the assumptions that the star is spherical (i.e. its centrifugal deformation is neglected) and uniformly rotating. However, it has recently been generalised to include the effects of a moderate centrifugal deformation using a perturbative approach.}
{We study the feasibility of carrying out a new generalisation to account for the centrifugal acceleration in the case of strongly deformed uniformly and rapidly rotating stars (and planets), and to identify the validity domain of this approximation.}
{We built a complete formalism analytically that allows the study of the dynamics of GIWs in spheroidal coordinates which take the flattening of uniformly and rapidly rotating stars into account by assuming the hierarchies of frequencies adopted within the TAR in the spherical case.}
{Using $2$D stellar models, we determine the validity domain of the generalised TAR as a function of the rotation rate of the star normalised by its critical angular velocity and its pseudo-radius. Assuming the anelastic and the two-dimensional Jeffreys-Wentzel-Kramers-Brillouin (JWKB) approximations, we derive a generalised Laplace tidal equation for the horizontal eigenfunctions of the GIWs and their asymptotic wave periods, which can be used  to probe the structure and dynamics of rotating deformed stars with asteroseismology.
The generalised TAR  where the centrifugal deformation of a star (or planet) is taken into account non-perturbatively allows us to identify, within the framework of 2D Evolution STEllaire en Rotation (ESTER) models, the validity domain of this approximation which is reduced by increasing the rate of rotation. We can affirm with a level of confidence of $90\%$ that the TAR remains applicable in all the space domain of deformed stars rotating at a rotation rate lower than $20\%$ of the critical rotation rate.}
{A new generalisation of the TAR, which takes the centrifugal acceleration into account in a non-perturbative way, is derived. This generalisation allows us to study the detectability and the signature of the centrifugal effects on GIWs in rapidly rotating deformed stars (and planets). We found that the effects of the centrifugal acceleration in rapidly rotating early-type stars on GIWs are theoretically detectable in modern space photometry using observations from \textit{Kepler}. We found also, by comparing the period spacing pattern computed with the standard and the generalised TAR, that the centrifugal acceleration affects the period spacing by increasing its values for low radial orders and by decreasing them slightly for high radial orders.}

\begin{document}
\maketitle

\section{Introduction}
The traditional approximation of rotation (TAR) was introduced for the first time in geophysics by \citet{eckart1960} to study the dynamics of the shallow Earth atmosphere and oceans. Then, \citet{Berthomieu1978} used it for the first time in an astrophysical context. Afterwards, \citet{Bildsten1996} and \citet{lee+saio1997} applied it to stellar pulsations to study low-frequency nonradial oscillations in rotating stars, more precisely the gravito-inertial waves (GIWs). These waves (often called $\mathrm{g}$ modes) propagate under the combined action of the buoyancy force and the Coriolis acceleration. A subclass of these waves are the Rossby waves (often called $\mathrm{r}$ modes) whose dynamics are driven by the conservation of the vorticity of the fluid.

The TAR is  applicable for low-frequency waves propagating in strongly stratified zones. In this case,  a hierarchy of frequencies is verified. Namely, the buoyancy force is stronger than the Coriolis force (i.e. $ 2 \Omega \ll N $, where $\Omega$ is the angular velocity and $N$ is the Brunt-Vaïsälä frequency)  in the direction of stable entropy or chemical stratification and the Brunt-Vaïsälä frequency is larger than the frequency of the waves in the co-rotating frame $\omega$ (i.e. $ \omega \ll N $). Considering these two assumptions, we can neglect the horizontal projection of the rotation vector $\Omega_{\mathrm{H}}=\Omega \sin \theta$ in the linearised hydrodynamic equations, where $\theta$ is the colatitude. Then, the vertical component of the Coriolis acceleration can be neglected and the wave velocity is almost   horizontal within the description of the dynamics of GIWs. In this framework, the linearised hydrodynamic system, which is coupled in the general case \citep{Dintrans1999, Dintrans+Rieutord2000}, becomes separable as in the non-rotating case and can be rewritten in the form of the Laplace tidal equation \citep{laplace1799}.

When the GIWs are observed with high-precision photometry, they allow us to study the internal structure and dynamics of stars with asteroseismology \citep{aerts2010, aerts2021}. GIWs also trigger transport of angular momentum and mixing inside the stars, modifying their evolution \citep[e.g.][]{mathis2008, mathis2009, rogers2013, rogers2015, aerts2019, Neiner2020}. In this context, GIWs which propagate in stably stratified stellar radiation regions are of major interest. First, they are detected and analysed with asteroseismology in intermediate- and high-mass stars \citep[e.g.][]{neiner2012, vanreeth2015} allowing the characterisation of the internal stellar rotation \citep{vanreeth2016, vanreeth2018}. In addition, GIWs are one of the candidates to explain the small rotation contrast observed between the stellar surface and core \citep {mathis2009, rogers2015, aerts2019}. So, the TAR and its flexibility allow us to derive powerful seismic diagnostics, based on the period spacings between consecutive radial order $\mathrm{g}$ modes in uniformly and differentially rotating spherical stars \citep{bouabid2013, ouazzani2017, vanreeth2018}. This enables us to determine the properties of the chemical stratification and  the rotation rate near the convective core of  early-type stars, which constrain our modelling of stellar structure, evolution and internal angular momentum transport \citep[e.g.][]{pedersen2018, aerts2018, aerts2019, ouazzani2019}.

In addition to the two conditions described above, three other assumptions are made when applying the TAR. First, the rotation is assumed to be uniform. Second, the star is assumed to be spherical, in other words, the centrifugal deformation of the star is neglected (i.e. $\Omega \ll \Omega_{\mathrm{K}} \equiv \sqrt{G M / R^{3}}$,  where $\Omega_{\mathrm{K}}$ is the Keplerian critical (breakup) angular velocity, and $G$, $M$, and $R$ are  the universal constant of gravity, the mass of the star, and the stellar radius, respectively). Third, the Cowling approximation \citep{cowling1941} is assumed where the fluctuations of the gravitational potential caused by the redistribution of mass by waves motion is neglected. 

In addition, the anelastic approximation \citep{Spiegel+Veronis1960} is assumed where the high-frequency acoustic modes can be filtered out. Finally, for asymptotic low frequency modes which are rapidly oscillating in the vertical direction, we can assume the JWKB (Jeffreys, Wentzel, Kramers, Brillouin) approximation \citep[e.g.][]{froman1965}.

The assumption of uniform rotation was abandoned by \citet{ogilvie+lin2004} and \citet{mathis2009}, who took into account the effects of general differential rotation (both in radius and latitude in a spherical star) on low-frequency GIWs within the framework of the TAR.  The results of this new formalism have been successfully applied in \citet{vanreeth2018} to derive the variation of the asymptotic period spacing in the case of a weak radial differential rotation and evaluate the sensitivity of GIWs to the effect of differential rotation.

The assumption of spherical symmetry was partially abandoned by \citet{mathis+prat2019}, who generalised the TAR for slightly deformed moderately rotating stars by considering the effects of the centrifugal acceleration using a perturbative approach. First, they derived the centrifugal perturbation of the spherical hydrostatic balance by determining the deformation of an isobar in the case of moderate uniform rotation. Afterwards, using this deformation they defined a mapping between the spherical coordinates system and a spheroidal coordinates system which takes into account this weak centrifugal perturbation of the star. Then, by keeping only first-order terms in the deformation, they re-derived the linearised hydrodynamic equations \citep{saio1981,lee+baraffe1995} and deduced the perturbed Laplace tidal equation within this new coordinates system. Then, they carried out a numerical exploration of the eigenvalues and horizontal eigenfunctions (the so-called Hough functions \citep{hough1898}) of the perturbed Laplace tidal equation. Finally, they derived the asymptotic expression for the frequencies of GIWs, including the effect of the centrifugal acceleration. This perturbative framework was then optimised for practical intense seismic forward modelling \citep{aerts2021} using one-dimensional stellar structure models by \cite{Henneco2021}.

In this work, we consider the general case of rapidly and uniformly rotating stars and planets which will generalise for the first time the previous work of \cite{lee+saio1997} and \cite{mathis+prat2019}. The shape of such objects becomes a strongly deformed spheroid because of the action of the centrifugal acceleration which cannot be modelled as a perturbation in this case. We study if the TAR could be generalised to include the effects of this deformation for any stellar (planetary) rotation including those close to their break-up velocity. The derived formalism could have several key applications. For instance, it could be used to build new seismic diagnostic and to study angular momentum transport \citep[e.g.][]{lee+saio1993,mathis2008,mathis2009,lee2014} and tidal dissipation \citep[e.g.][]{ogilvie+lin2004,ogilvie+lin2007,braviner2014} induced by low-frequency GIWs in deformed stars and planets.

To do so, we first introduce in Sect.\,\ref{sect:Hydrodynamic_equations} a new spheroidal coordinate system which follows the shape of a deformed star for any rotation. It is based on the work by \cite{bonazzola1998} and was first used by \cite{rieutord2005} to develop rotating polytropic models. Then  \citet{lignieres2006} employed it to study acoustic oscillations in strongly deformed rotating stars. Afterwards, \cite{reese2006,reese2009} and \cite{Ouazzani2012} make use of it to develop the two-dimensional oscillation program (TOP) and the adiabatic code of oscillation including rotation (ACOR), respectively. They compute non-radial pulsations of rotating stars. Finally, \citet{espinosa+rieutord2013} utilised it to develop the Evolution STEllaire en Rotation (ESTER) code which computes the structure of an early-type star including its differential rotation and the associated meridional circulation. Here, we first consider the case of uniform rotation to distinguish the effects of the deformation from those of differential rotation associated to the centrifugal and Coriolis accelerations, respectively. From there, we derive the system of linearised hydrodynamic equations in spheroidal coordinates. In Sect.\,\ref{sect:Generalised_TAR}, we apply the TAR in this new coordinate system by adopting the adequate assumptions. In Sect.\,\ref{sect:Dynamics_GIWs}, we can then rewrite the oscillation equations in the form of a new generalised Laplace tidal equation (GLTE) and deduce the asymptotic frequencies of low-frequency GIWs. In Sect.\,\ref{sect:results}, we identify the validity domain of the TAR when using ESTER models and carry out a  numerical exploration of the eigenvalues and the eigenfunctions of the GLTE. In Sect.\,\ref{sect:seismic_diagnosis}, we quantify the centrifugal deformation effect of the star on the asymptotic period spacing pattern and we discuss its detectability. In Sect.\,\ref{sect:hierarchy_validation}, we verify the validity of the hierarchy of the terms in the linearised hydrodynamics  equations imposed by the TAR in deformed stars. Finally, we discuss and summarise our work and results in Sect.\,\ref{sect:conclusion}.

\section{Hydrodynamic equations in uniformly rotating deformed stars and planets} \label{sect:Hydrodynamic_equations}
\subsection{Spheroidal geometry}
\begin{figure}
    \centering
    \resizebox{\hsize}{!}{\includegraphics{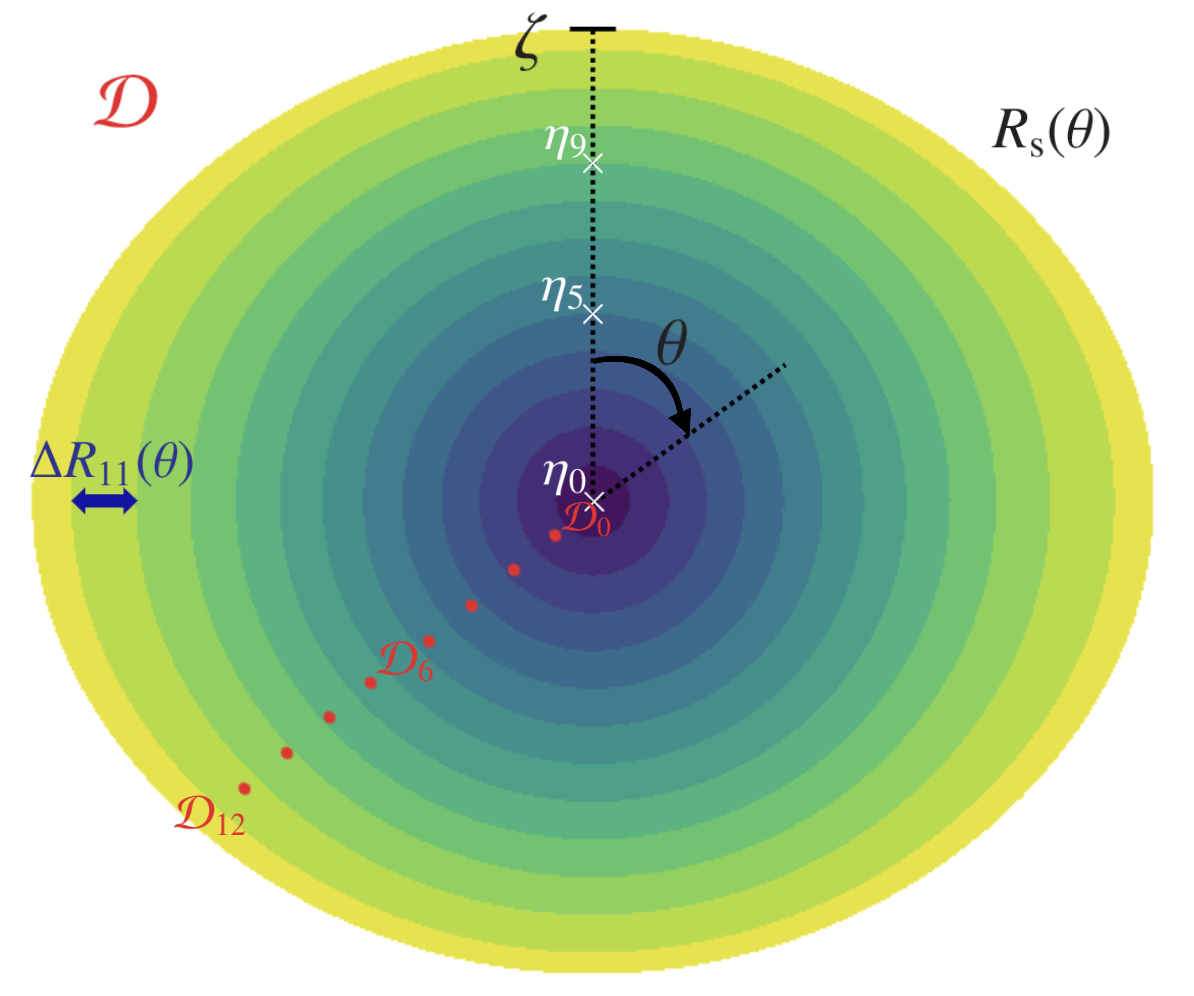}}
    \caption{Spheroidal coordinate system with $n=13$ subdomains used to compute the equilibrium model of a star rotating at $60\%$ of its break-up velocity corresponding to the case of the $3 \mathrm{M}_{\odot}$ ESTER model (with the central fraction in Hydrogen $X_{\rm c} = 0.7$) studied in Sect.\,\ref{sect:results}. $\zeta$ and $\theta$ are the pseudo-radius and the colatitude, respectively. $\eta_i$ and $\Delta R_i(\theta)$ are defined in our mapping (Eq.\;\ref{eq:mapping}).}
    \label{fig:spheroidal_geometry}
\end{figure}

Because of the flattened shape of a rapidly rotating star, we use the spheroidal coordinate system  $(\zeta, \theta, \varphi)$ proposed by \citet{bonazzola1998}, where $\zeta$ is the radial coordinate (the pseudo-radius), $\theta$ the colatitude and $\varphi$ the azimuth. It was designed not only to fit the shape of the star at the surface but also to get closer to the spherical shape at the centre as illustrated in Fig.\,\ref{fig:spheroidal_geometry}. Following \citet{rieutord+espinosa2013}, this new coordinate system can be linked to the usual spherical one $(r, \theta, \varphi)$ via the following mapping
\begin{equation}\label{eq:mapping}
    r(\zeta, \theta)=a_{i} \xi \Delta \eta_{i}+R_{i}(\theta)+A_{i}(\xi)\left(\Delta R_{i}(\theta)-a_{i} \Delta \eta_{i}\right), \text { for } \zeta \in\left[\eta_{i}, \eta_{i+1}\right],
\end{equation}
where we split the spheroidal domain $\mathcal{D}$ in $n$ subdomains $\mathcal{D}_{i \in  \llbracket0,n-1 \rrbracket} \in\left[R_{i}(\theta), R_{i+1}(\theta)\right] $ where $R_{i \in  \llbracket0,n \rrbracket}(\theta)$ are series of functions, such that $R_{n}(\theta)=R_{\rm s}(\theta)$ is the outer boundary and $R_{0}(\theta)=0$ is the centre. Additionally, $\eta_{i} =R_{i}(\theta=0)$ are the polar radii of the interfaces between the subdomains, $\Delta \eta_{i} =\eta_{i+1}-\eta_{i}$, $\Delta R_{i}(\theta) =R_{i+1}(\theta)-R_{i}(\theta)$ and $\xi =(\zeta-\eta_{i})/\Delta \eta_{i}$. The functions $A_{i \in  \llbracket1,n-1 \rrbracket}(\xi)=-2 \xi^{3}+3 \xi^{2}$ and $A_{0}(\xi)=-1.5 \xi^{5}+2.5 \xi^{3}$ and the constants $a_{i}$ (in practice $a_{i}=1$) are chosen to satisfy the boundary conditions between the different subdomains so that the mapping is continuous and derivable.
The other coordinates ($\theta$ and $\varphi$) remain unchanged. Since we choose the polar radius as the unit length $R_{\rm s}(\theta=0) = 1$, $\zeta=1$ coincides with the surface (i.e. $r(\zeta = 1, \theta)=R_{\rm s}(\theta)$), while $r=0$ is obtained at the centre when $\zeta=0$ (i.e. $r(\zeta = 0, \theta)=0$). Along the rotation axis $(\theta=0)$, the deformed radial coordinate $\zeta$ is the distance from the centre normalised by the polar radius and when the centrifugal deformation is zero, $\zeta$ is equivalent to the radius, hence the name pseudo-radius. 

Once the spheroidal coordinate system is established, it is mandatory to specify the corresponding basis. To do so, we define the spheroidal basis $(\vec{a}_{\zeta}, \vec{a}_{\theta}, \vec{a} _ {\varphi})$ from the natural covariant basis $(\vec{E}_{\zeta}, \vec {E}_{\theta}, \vec{E}_{\varphi})$ defined as $\vec{E}_{i}=\partial_{i} (r\vec{e_r})$
\begin{equation}\label{eq:spheroidal_base}\left\{\begin{array}{lcc} \begin{aligned}
\vec{a}_{\zeta} &=\frac{\zeta^{2}}{r^{2} r_{\zeta}} \vec{E}_{\zeta}=\frac{\zeta^{2}}{r^{2}} \vec{e}_{r} \\
\vec{a}_{\theta} &=\frac{\zeta}{r^{2} r_{\zeta}} \vec{E}_{\theta}=\frac{\zeta}{r^{2} r_{\zeta}}\left(r_{\theta} \vec{e}_{r}+r \vec{e}_{\theta}\right) \\
\vec{a}_{\varphi} &=\frac{\zeta}{r^{2} r_{\zeta} \sin \theta} \vec{E}_{\varphi}=\frac{\zeta}{r r_{\zeta}} \vec{e}_{\varphi}
\end{aligned}\end{array}\right.,\end{equation}
where $r_\zeta\equiv\partial_\zeta r$, $r_\theta\equiv\partial_\theta r$ and $(\vec{e}_{r}, \vec{e}_{\theta}, \vec{e} _ {\varphi})$ is the usual spherical unit-vector basis. This basis has the advantage of being reduced to the usual spherical unit-vector basis when we approach the spherical geometry ($R_{\rm s}(\theta)$ tends to the constant function 1).

\subsection{Linearised hydrodynamic equations in spheroidal coordinates}
Once the geometry is fixed, we can derive  the complete adiabatic inviscid system of linearised equations to treat the wave dynamics in a uniformly rotating, strongly deformed star or planet. First, the linearised momentum equation for an inviscid fluid is written as (we refer the reader to Appendix\;\ref{app:momentum_equation} for a detailed derivation)
\begin{multline} \label{eq:radial_momentum}
    (\partial_t+\Omega\partial_\varphi) \left[\frac{\zeta^2 r_{\zeta} }{r^{2}}v^{\zeta}+\frac{\zeta r_{\theta} }{r^{2}}v^{\theta}\right]
    \\=2 \Omega     \frac{\zeta \sin \theta }{r}v^{\varphi}-\frac{1}{\rho_{0}}\partial_{\zeta} P^{\prime}+\frac{\rho^{\prime}}{\rho_{0}^2} \partial_{\zeta} P_{0}- \partial_{\zeta} \Phi^{\prime},
\end{multline}

\begin{multline} \label{eq:latitudinal_momentum}
    (\partial_t+\Omega\partial_\varphi)  \left[\frac{\zeta^2 r_{\theta}}{r^{2}}v^{\zeta} + \frac{\zeta \left( r^{2}+r_{\theta}^{2} \right)}{r^{2} r_{\zeta}}v^{\theta}\right]\\
    =2 \Omega    \frac{\zeta\left(r_{\theta} \sin \theta+r \cos \theta \right) }{r r_{\zeta}}v^{\varphi}-\frac{1}{\rho_{0}}\partial_{\theta} P^{\prime}+\frac{\rho^{\prime}}{\rho_{0}^2} \partial_{\theta} P_{0}- \partial_{\theta} \Phi^{\prime},
\end{multline}

\begin{multline} \label{eq:azimuthal_momentum}
    \frac{\zeta }{r_{\zeta}}(\partial_t+\Omega\partial_\varphi) v^{\varphi}=- 2 \Omega \frac{ \zeta^{2} \sin \theta }{r} v^{\zeta}
    \\-2 \Omega \frac{ \zeta\left(r_{\theta} \sin \theta+r \cos \theta\right) }{r r_{\zeta}}v^{\theta}-\frac{1}{\rho_0 \sin \theta}\partial_{\varphi} P^{\prime}-\frac{1}{\sin \theta} \partial_{\varphi} \Phi^{\prime},
\end{multline}
where $v^{\zeta}$, $v^{\theta}$ and $v^{\varphi}$ are the contravariant components of the velocity field $\vec{v}=v^{i} \vec{a}_{i}$ and $\vec{\Omega}=\Omega(\cos{\theta}\vec{e_r}-\sin{\theta}\vec{e_\theta})$ is the uniform angular velocity. $\rho$, $\Phi$ and $P$ are the fluid density, gravitational potential and pressure, respectively. Each of these scalar quantities has been expanded as
\begin{equation*}
X(\zeta, \theta, \varphi, t)=X_0(\zeta,\theta)+X^\prime(\zeta, \theta, \varphi, t),
\end{equation*}
where $X_0$ is the hydrostatic component of $X$ and $X^\prime$ the wave's associated linear fluctuation. When assuming the \citet{cowling1941} approximation, the fluctuation of the gravitational potential is neglected ($\Phi^\prime = 0$).

Subsequently, the linearised continuity equation is obtained (we refer the reader to  Appendix\;\ref{app:continuity_equation} for a detailed derivation)
\begin{multline} \label{eq:continuity}
    (\partial_t+\Omega\partial_\varphi)  \rho^{\prime}=-\frac{\zeta^{2} \partial_{\zeta} \rho_{0}}{r^{2} r_{\zeta}} v^{\zeta}-\frac{\zeta \partial_{\theta} \rho_{0}}{r^{2} r_{\zeta}} v^{\theta}
    \\-\frac{\zeta^{2} \rho_{0}}{r^{2} r_{\zeta}}\left[\frac{\partial_{\zeta}\left(\zeta^{2} v^{\zeta}\right)}{\zeta^{2}}+\frac{\partial_{\theta}\left(\sin \theta v^{\theta}\right)}{\zeta \sin \theta}+\frac{\partial_{\varphi} v^{\varphi}}{\zeta \sin \theta}\right].
\end{multline}

Then, the linearised energy equation in the adiabatic limit is derived (we focus in this work on adiabatic oscillations as in studies of the TAR in spherical stars)
\begin{equation}\label{eq:energy_eq}
    (\partial_t+\Omega\partial_\varphi) \left(\frac{1}{\Gamma_{1}}\frac{P^{\prime}}{P_0}- \frac{\rho^{\prime}}{\rho_0}\right)=\frac{ N^{2} }{\left\|\vec{g}_{\rm eff}\right\|^{2}} \vec{v} \cdot \vec{g}_{\rm eff},
\end{equation}
where $\Gamma_{1}=(\partial \ln P_0 / \partial \ln \rho_0)_{S}$ ($S$ being the macroscopic entropy) is the adiabatic exponent, $\vec{g}_{\rm eff}=-\vec{\nabla} \Phi_0+\frac{1}{2}\Omega^2\vec{\nabla}(r^2\sin^2{\theta})=\vec{\nabla}P_0/\rho_0$ is the background effective gravity which includes the centrifugal acceleration and $N^2$ is the squared Brunt–Väisälä frequency given by
\begin{equation}
    N^{2}(\zeta, \theta)=\vec{g}_{\rm eff} \cdot\left(-\frac{1}{\Gamma_{1}} \frac{\boldsymbol{\nabla} P_{0}}{P_{0}}+\frac{\boldsymbol{\nabla} \rho_{0}}{\rho_{0}}\right).
\end{equation}

Finally, since in this work we are interested in studying standing eigenmodes of oscillation we consider a discrete spectrum of eigenfrequencies rather than an integral over a continuous spectrum. So, we expand the wave's velocity and scalar quantities fluctuations $(X^\prime \equiv \{\rho^\prime, P^\prime\}$) on discrete Fourier series both in time and in azimuth
\begin{gather}
    \vec{v}(\zeta, \theta, \varphi, t)  \equiv \sum_{\omega^{\mathrm{in}}, m}\left\{\vec{u}(\zeta, \theta) \exp [i(\omega^{\mathrm{in}} t-m \varphi)]\right\},\\
    X^{\prime}(\zeta, \theta, \varphi, t)  \equiv \sum_{\omega^{\mathrm{in}}, m}\left\{\widetilde{X}(\zeta, \theta) \exp [i(\omega^{\mathrm{in}} t-m \varphi)]\right\},
\end{gather}
where $m$ is the azimutal order ($m$ is an integer because we are looking for a linear solution of a wave problem whose axisymmetric background depends only on $\zeta$ and $\theta$) and $\omega^{\mathrm{in}}$ is the wave eigenfrequency in an inertial reference frame (we don't put an index for the eigenfrequencies in order to make the notations simpler). In a uniformly rotating region, the waves are Doppler shifted due to the rotation. We can define the wave eigenfrequency $\omega$ in the co-rotating reference frame  as
\begin{equation} \label{eq:doppler_shift}
    \omega = \omega^{\mathrm{in}} - m\Omega.
\end{equation}

\section{Generalised TAR}\label{sect:Generalised_TAR}
\subsection{Approximation  on the stratification profile: $N^2(\zeta, \theta)\approx N^2(\zeta)$} \label{subsect:hypothesis1}
To obtain a separable system of equations by applying the TAR, a first necessary condition is that the Brunt-Väisälä frequency $N^2$  depends mainly on the pseudo-radius $\zeta$, that is to say it has a small variation with respect to the colatitude $\theta$. This also implies that the hydrostatic pressure $P_0$ and the hydrostatic density $\rho_0$ depend mainly on $\zeta$ so the linearised energy equation simplifies into
\begin{equation}\label{eq:energy_eq_simplified}
     \frac{1}{\Gamma_{1}}\frac{\widetilde{P}}{P_0}- \frac{\widetilde{\rho}}{\rho_0}=\frac{1}{i\omega}\frac{ N^{2} }{g_{\rm eff}} \frac{\zeta^2}{r^2r_\zeta}u^\zeta.
\end{equation}

In order to compute the relative error made by adopting this approximation and to define its validity domain, we compare in Sect.\,\ref{subsect:validity} an approximate value of $N^2$ where it depends only on $\zeta$ to the exact one depending on $\zeta$ and $\theta$ by fixing the value of $\theta$ and varying the angular velocity of the star using two-dimensional ESTER stellar models.

\subsection{The TAR with centrifugal acceleration}\label{subsec:tar}

Using this first approximation, and assuming the same hierarchies of frequencies ($ 2\Omega\ll N$ and $\omega\ll N$) and velocity scales ($|v^\zeta|\ll\{|v^\theta|,|v^\varphi|\}$) that in the spherical and in the weakly deformed cases (this will be discussed in details in Sect.\;\ref{sect:hierarchy_validation}), we now built the generalised framework for the TAR in the case of a uniformly and rapidly rotating strongly deformed star (planet).

By adopting the approximation $N^2(\zeta, \theta)\approx N^2(\zeta)$ and the Cowling approximation, we can rewrite the radial momentum equation (\ref{eq:radial_momentum}) as
\begin{multline} \label{eq:radial_momentum2}
    i\omega \left[\frac{\zeta^2 r_{\zeta} }{r^{2}}u^{\zeta}+\frac{\zeta r_{\theta} }{r^{2}}u^{\theta}\right] \\ 
    = 2 \Omega     \frac{\zeta \sin \theta }{r}u^{\varphi} -\partial_{\zeta} \widetilde{W} - \frac{\widetilde{P}}{\rho_0^2}\partial_\zeta \rho_0 - \frac{N^2 }{i\omega}\frac{\zeta^2}{r^2 r_\zeta} u^\zeta,
\end{multline}
where $\widetilde{W}=\widetilde{P}/\rho_0$ is the normalised pressure. The propagation of low-frequency GIWs can be studied within the anelastic approximation in which acoustic waves are filtered out \citep{Dintrans+Rieutord2000, mathis2009}. Therefore, Eqs.\;(\ref{eq:radial_momentum2}) and (\ref{eq:energy_eq_simplified}) can be simplified accordingly by neglecting the terms $(\widetilde{P} / \rho_0^{2}) \partial_{\zeta} \rho_0 $ and $(1 / \Gamma_{1}) \widetilde{P} / P_0$. Within the TAR, we focus on low-frequency waves (i.e. $\omega \ll N$) propagating in strongly stratified regions (i.e. $2\Omega \ll N$). In this case, we can neglect the vertical component of the Coriolis acceleration because it is dominated by the buoyancy force and the vertical wave velocity since $|v^\zeta|\ll\{|v^\theta|,|v^\varphi|\}$ because of the strong stable stratification. Thus, the radial momentum equation can be simplified into
\begin{equation} \label{eq:radial_momentum_TAR}
    i{\omega} \partial_\zeta \widetilde{W}+N^2 \zeta^2 \mathcal{A} u^\zeta=0,
\end{equation}
where
\begin{equation}
    \mathcal{A}(\zeta,\theta)=\frac{1}{r^2 r_\zeta}.
\end{equation}
Subsequently, we examine the latitudinal component of the momentum
equation (Eq\;\ref{eq:latitudinal_momentum}), which reduces, for the same reasons, to
\begin{equation} \label{eq:latitudinal_momentum_TAR}
    i{\omega} \zeta \mathcal{B} u^\theta-2 \Omega \zeta \mathcal{C}u^\varphi+\partial_\theta \widetilde{W}=0,
\end{equation}
where
\begin{gather}
    \mathcal{B}(\zeta,\theta)=\frac{r^2+r_\theta^2}{r^2r_\zeta},
    \\\mathcal{C}(\zeta,\theta)=\frac{r_\theta\sin{\theta+r \cos{\theta}}}{r r_\zeta}.
\end{gather}
Finally, the azimuthal component of the momentum equation (Eq.\;\ref{eq:azimuthal_momentum}) simplifies onto
\begin{equation} \label{eq:azimuthal_momentum_TAR}
    i{\omega} \zeta \mathcal{D} u^\varphi+2\Omega\zeta \mathcal{C}u^\theta-\frac{i m}{\sin{\theta}} \widetilde{W}=0,
\end{equation}
where
\begin{equation}
    \mathcal{D}(\zeta,\theta)=\frac{1}{r_\zeta}.
\end{equation}
We thus obtain a system of equations for strongly deformed rotating stars which has the same mathematical form that in the case of spherically symmetric stars \citep{lee+saio1997} and weakly deformed stars \citep{mathis+prat2019}, and where we thus still manage to partially decouple the vertical and horizontal components of the velocity. By solving the system formed by Eqs.\;(\ref{eq:latitudinal_momentum_TAR}) and (\ref{eq:azimuthal_momentum_TAR}) we can express $u^\theta$ and $u^\varphi$ as a function of $\widetilde{W}$ as follows
\begin{multline}\label{eq:u_theta(zeta,theta)}
    u^\theta(\zeta, \theta)=i\frac{1}{\omega}\frac{1}{\zeta}\frac{1}{\mathcal{B}(\zeta, \theta)}\\\left[\left(1+\nu^2\frac{\mathcal{C}^2(\zeta, \theta)}{\mathcal{E}(\zeta, \theta)\mathcal{B}(\zeta, \theta)}\right)\partial_\theta \widetilde{W}-\frac{m\nu}{\sin{\theta}}\frac{  \mathcal{C}(\zeta, \theta)}{\mathcal{E}(\zeta, \theta)}\widetilde{W}\right],
\end{multline}
\begin{equation} \label{eq:u_phi(zeta,theta)}
    u^\varphi(\zeta, \theta)=-\frac{1}{\omega}\frac{1}{\zeta}\frac{1}{\mathcal{E}(\zeta, \theta)}\left[\nu\frac{ \mathcal{C}(\zeta, \theta)}{\mathcal{B}(\zeta, \theta)}\partial_\theta \widetilde{W}- \frac{m}{\sin{\theta}}\widetilde{W} \right],
\end{equation}
where we introduce
\begin{equation}
    \mathcal{E}(\zeta,\theta)=\mathcal{D}(\zeta,\theta)-\nu^2\frac{\mathcal{C}^2(\zeta,\theta)}{\mathcal{B}(\zeta,\theta)},
\end{equation}
and the spin parameter
\begin{equation}
    \nu=\frac{2 \Omega}{\omega}.
\end{equation}
The structure of the new equations that we obtain in the spheroidal case is similar to the one in the usual spherical case \citep[e.g.][]{lee+saio1997,townsend2003,pantillon2007}. The most important difference is that the coefficients $\mathcal{A}$, $\mathcal{B}$, $\mathcal{C}$, $\mathcal{D}$ and $\mathcal{E}$ are function of $\zeta$ and $\theta$ through $r(\zeta,\theta)$ and their derivatives. As this is done in \citet{lee+saio1997} and \citet{mathis+prat2019}, we introduce
the reduced latitudinal coordinate $x=\cos{\theta}$, so that Eqs.\;(\ref{eq:u_theta(zeta,theta)}) and (\ref{eq:u_phi(zeta,theta)}) transform to

\begin{align}\label{eq:u_theta(zeta,x)}
    u^\theta(\zeta, x&)=\mathcal{L}_{\nu m}^{\theta}\left[\widetilde{W}(\zeta, x)\right] \nonumber\\
    \begin{split}
    &\mathrel{\phantom{)}}=-i\frac{1}{\omega}\frac{1}{\zeta}\frac{1}{\mathcal{B}(\zeta, x)}\frac{1}{\sqrt{1-x^2}}\\
    &\left[\left(1-x^2\right)\left(1+\nu^2\frac{\mathcal{C}^2(\zeta, x)}{\mathcal{E}(\zeta, x)\mathcal{B}(\zeta, x)}\right)\partial_x +m\nu\frac{  \mathcal{C}(\zeta, x)}{\mathcal{E}(\zeta, x)}\right]\widetilde{W},
    \end{split}
\end{align}

\begin{align} \label{eq:u_phi(zeta,x)}
    u^\varphi(\zeta, x) &= \mathcal{L}_{\nu m}^{\varphi}\left[\widetilde{W}(\zeta, x)\right] \nonumber\\
    &=\frac{1}{\omega}\frac{1}{\zeta}\frac{1}{\mathcal{E}(\zeta, x)}\frac{1}{\sqrt{1-x^2}}\left[\left(1-x^2\right)\nu\frac{ \mathcal{C}(\zeta, x)}{\mathcal{B}(\zeta, x)}\partial_x + m \right]\widetilde{W}.
\end{align}

\section{Dynamics of low-frequency gravito-inertial waves}\label{sect:Dynamics_GIWs}
Our goal in this section is to derive the generalised Laplace tidal equation (GLTE) for the normalised pressure $\widetilde{W}$, which allows us to compute the frequencies and periods of low-frequency GIWs and to build the corresponding seismic diagnostics following the method by \cite{bouabid2013}, \cite{vanreeth2018}, and \cite{mathis+prat2019}. Applying the anelastic approximation and the approximation  $N^2(\zeta, \theta)\approx N^2(\zeta)$ in the continuity equation (Eq.\;\ref{eq:continuity}) allows us to neglect the terms $\partial_t \rho^\prime$ and $\partial_\theta \rho_0$ respectively, so it simplifies into 
\begin{equation}\label{eq:continuity_simplified}
    \zeta \partial_\zeta \rho_0 u^\zeta+\rho_0\left[\frac{\partial_{\zeta}\left(\zeta^{2} u^{\zeta}\right)}{\zeta}+\frac{1}{\sin{\theta}}\partial_\theta(\sin{\theta}u^\theta)-\frac{i m }{\sin{\theta}}u^\varphi\right]=0.
\end{equation}

\subsection{JWKB approximation}
Assuming that $\omega\ll N$, each scalar field and each component of $\vec{u}$ can be expanded using the two-dimensional Jeffreys-Wentzel-Kramers-Brillouin (JWKB) approximation which was introduced for strongly stratified differentially rotating fluids by \citet{mathis2009}. In this case, the vertical wave number $k_V$ ($\equiv k_\zeta$) is assumed to be very large, and the associated wavelength is thus very small. Therefore, the spatial variation of the wave in the pseudo-radial direction is very fast compared to that of the hydrostatic background given by $\rho_0 $, $P_0$ and $ g_{\rm eff}$. 
So the spatial structure of the waves can be described by the product of a rapidly oscillating plane-like wave function in the pseudo-radial direction multiplied by a bi-dimensional slowly varying envelope:
\begin{align}\label{eq:jwkb_w}
\widetilde{W}(\zeta, \theta)&=\sum_{k}\left\{w_{\nu k m}(\zeta, \theta) \frac{A_{\nu k m}}{k_{V ; \nu k m}^{1 / 2}} \exp \left[i \int^{\zeta} k_{V ; \nu k m} \mathrm{d} \zeta\right]\right\},
\\u^{j}(\zeta, \theta)&=\sum_{k}\left\{\hat{u}^j_{\nu k m}(\zeta, \theta) \frac{A_{\nu k m}}{k_{V ; \nu k m}^{1 / 2}} \exp \left[i \int^{\zeta} k_{V;\nu k m} \mathrm{d} \zeta \right]\right\},\label{eq:jwkb_u}
\end{align}
with $j \equiv\{\zeta, \theta, \varphi\}$, $k$ is the index of a latitudinal eigenmode (cf. Sect.\;\ref{subsect:glte}) and $A_{\nu k m}$ is the amplitude of the wave. Substituting the expansion given in Eqs.\;(\ref{eq:jwkb_w}) and (\ref{eq:jwkb_u}) into Eqs.\;(\ref{eq:radial_momentum_TAR}), (\ref{eq:u_theta(zeta,x)}) and (\ref{eq:u_phi(zeta,x)}), the final pseudo-radial, latitudinal and azimuthal components of the velocity are obtained:
\begin{align}
    \hat{u}^\zeta_{\nu k m}(\zeta, x) &=  \frac{ k_{V;\nu k m}(\zeta)}{N^2(\zeta)} \frac{\omega}{\zeta^2 \mathcal{A}(\zeta, x)} w_{\nu k m}(\zeta, x), \label{eq:u_zeta_final}
    \\\hat{u}^\theta_{\nu k m}(\zeta, x) &= \mathcal{L}_{\nu m}^{\theta}\left[w_{\nu k m}(\zeta, x)\right], \label{eq:u_theta_final}
    \\\hat{u}^\varphi_{\nu k m}(\zeta, x) &= \mathcal{L}_{\nu m}^{\varphi}\left[w_{\nu k m}(\zeta, x)\right].\label{eq:u_phi_final}
\end{align}

\subsection{Approximation: $\mathcal{A}(\zeta, \theta)\approx \mathcal{A}(\zeta)$} \label{subsubsect:hypothesis2}
To be able to derive a generalised Laplace tidal equation as in the case of slightly deformed spheres \citep{mathis+prat2019} and to introduce its horizontal eigenvalues $\Lambda_{\nu k m}$, which should depend only on $\zeta$, we have to do a partial separation between the pseudo-radial and latitudinal variables in the pseudo-radial component of the velocity $\hat{u}^\zeta$. In other words, we have to write Eq.\;(\ref{eq:u_zeta_final}) as the product of a function which depends only on $\zeta$ and the normalised pressure $w_{\nu k m}$. To do so, we have to assume that the coefficient $\mathcal{A}(\zeta,\theta)$ depends mainly on $\zeta$. To evaluate the relative error made by adopting this approximation and to define its validity domain, we compare in Sect.\,\ref{subsect:validity} an approximate value of $\mathcal{A}$ where it depends only on $\zeta$ to the exact one depending on $\zeta$ and $\theta$ by fixing the value of $\theta$ and varying the angular velocity of the star using two-dimensional ESTER stellar models.

\subsection{Generalised Laplace tidal equation (GLTE)} \label{subsect:glte}
Substituting the radial component of the velocity (Eq.\;\ref{eq:u_zeta_final}) into the continuity equation (Eq.\;\ref{eq:continuity_simplified}) we get:
\begin{equation}\label{eq:continuity_final}
        \frac{ \omega^2_{k m} k^2_{V;\nu k m}}{\mathcal{A} N^2}  w_{\nu k m}  + \underbrace{ \omega_{k m} \zeta \left(i\partial_x(\sqrt{1-x^2}\hat{u}^\theta)-\frac{m }{\sqrt{1-x^2}}\hat{u}^\varphi\right)}_{=\mathcal{L}_{\nu m}\left[w_{\nu k m}\right]}=0,
\end{equation}
where the JWKB approximation allows us to neglect  $\zeta \partial_\zeta \rho_0/\rho_0 u^\zeta$ and $2 u^\zeta$ in front of the dominant term $\zeta \partial_\zeta u^\zeta $.
Using Eqs. (\ref{eq:u_theta_final}) and (\ref{eq:u_phi_final}), we obtain the GLTE for the normalised pressure $w_{\nu k m}$
\begin{align}
    \begin{split}
    \mathcal{L}_{\nu m}\left[w_{\nu k m}\right]& = \partial_x\left[\frac{1}{\mathcal{B}}\left(1+\frac{\nu^2\mathcal{C}^2}{\mathcal{E}\mathcal{B}}\right)(1-x^2)\partial_x w_{\nu k m}\right]
    \\&\mathrel{\phantom{=}}+\left[m \nu \partial_x\left(\frac{ \mathcal{C}}{\mathcal{E}\mathcal{B}}\right) -\frac{m^2}{\mathcal{E}(1-x^2)}\right]w_{\nu k m}
    \end{split} \nonumber
    \\&= -\Lambda_{\nu k m}(\zeta) w_{\nu k m}.\label{eq:glte}
\end{align} 
When the centrifugal acceleration is taken into account, the eigenvalues $\Lambda_{\nu km}$ and the generalised Hough functions (eigenfunctions) $w_{\nu km}$ of the GLTE vary with the pseudo-radius $\zeta$. We choose to define our latitudinal ordering number $k$ to enumerate, for each $(\nu,m)$, the infinite set of solutions as in \citet{lee+saio1997} and \citet{mathis+prat2019} by considering the eigenvalues and eigenfunctions at the centre where they are not affected by the centrifugal acceleration since our mapping (Eq.\;\ref{eq:mapping}) is such that $r(\zeta \rightarrow 0,\theta) \rightarrow 0$.

We can check the validity of our model analytically by calculating the coefficients of the GLTE (Eq.\;\ref{eq:glte}) in the spherical case (cf.\;Table\;\ref{table:termes}) and by substituting these values into the GLTE. We then recover the standard Laplace tidal equation (SLTE) of \citet{lee+saio1997}:
\begin{align}
    \begin{split}
    \mathcal{L}^{\rm{stand.}}_{\nu m} \left[w^{\rm{stand.}}_{\nu k m}\right] & = \partial_{x}\left[ \frac{1-x^2}{1-\nu^2 x^2}\partial_{x} w^{\rm{stand.}}_{\nu k m} \right] 
    \\&\mathrel{\phantom{=}} + \left[m\nu\frac{1+\nu^2 x^2}{\left(1-\nu^2 x^2\right)^2} - \frac{m^2}{\left(1-x^2\right)\left(1-\nu^2 x^2\right)}\right]w^{\rm{stand.}}_{\nu k m}
    \end{split}\nonumber
    \\&= -\Lambda^{\rm{stand.}}_{\nu k m} w^{\rm{stand.}}_{\nu k m}.\label{eq:slte}
\end{align}
In this case, the eigenvalues $\Lambda^{\rm{stand.}}_{\nu k m}$ are independent of the pseudo-radius $\zeta$. We note that at the limit $\nu=0$, which corresponds to the non-rotating case, the SLTE reduces to the differential equation for the associated Legendre polynomials $P_{\ell, m}$ of degree $\ell$ and order $m$ with $\Lambda^{\rm{stand.}}_{0km}= \ell(\ell + 1)$ and $ \ w_{0km} (x) = a _{\ell, m} P_{\ell, m}(x)$ where $ \ell = | m | + k $ and 
\begin{equation*}
a _{\ell, m}=(-1)^{\frac{m+|m|}{2}}\displaystyle{\sqrt{\frac{2 l+1}{4 \pi} \frac{(l-|m|) !}{(l+|m|) !}}},
\end{equation*}
is the normalisation coefficient.

\begin{table*}
    \centering
    \caption{Terms involved in the derivation of the GLTE in the general case of spheroidal geometry and in the particular case of spherical geometry (we refer the reader to Appendix\;\ref{app:terms} for a visual presentation of these terms).}
    \label{table:termes}
        \begin{tabular}{c|c|c}
        \hline \hline
        Terms & Spheroidal geometry & Spherical geometry
        \\\hline \hline
        $\mathcal{A}$ & $\displaystyle{\frac{1}{r^2 r_\zeta}}$ & $\displaystyle{\frac{1}{\zeta^2}}$ \\\hline
        $\mathcal{B}$ & $ \displaystyle{\frac{r^2+(1-x^2)r_x^2}{r^2r_\zeta}}$ & $1$ 
        \\\hline
        $\mathcal{C}$ & $ \displaystyle{\frac{-(1-x^2) r_x+r x}{r r_\zeta}}$ & $x$ \\\hline
        $\mathcal{D}$ & $\displaystyle{\frac{1}{r_\zeta}}$ & $1$ 
        \\\hline
        $\mathcal{E}$ & $\displaystyle{\mathcal{D}-\nu^2\frac{\mathcal{C}^2}{\mathcal{B}}=\frac{1}{r_\zeta}-\nu^2 \frac{\left(-(1-x^2) r_x+r x\right)^2}{r^2+(1-x^2)r_x^2}}$ & $1-\nu^2x^2$ 
        
        \\\hline
        $ \partial_x \mathcal{B} $ & $\displaystyle{\frac{2 r r_x - 2 x r^2_x + 2 (1-x^2) r_x r_{xx} - \mathcal{B}\left(2 r r_x r_\zeta +r^2 r_{\zeta x}\right)}{r^2 r_\zeta}}$ & $0$
        \\\hline
        $ \partial_x \mathcal{C} $ & $\displaystyle{\frac{2x r_x -(1-x^2)r_{xx} + r+ x r_x - \mathcal{C}\left(r_x r_\zeta+ r r_{\zeta x}\right)}{r r_\zeta}}$ & $1$ 
        \\\hline
        $ \partial_x \mathcal{D} $ & $-\mathcal{D}^2 r_{\zeta x}$ & $0$ 
        \\\hline
        $\displaystyle{\partial_x\left(\frac{ \mathcal{C}}{\mathcal{E}\mathcal{B}}\right)}$& $   \displaystyle{\frac{\partial_x\mathcal{C}\left(\mathcal{D}\mathcal{B}-\nu^2r_\zeta\mathcal{C}^2\right)-\mathcal{C}\left(\mathcal{B}\partial_x\mathcal{D}+\mathcal{D}\partial_x\mathcal{B}-\nu^2r_{\zeta x}\mathcal{C}^2-2\nu^2r_\zeta \mathcal{C}\partial_x \mathcal{C}\right)}{\left(\mathcal{D}\mathcal{B}-\nu^2r_\zeta\mathcal{C}^2\right)^2}}$ & $\displaystyle{\frac{1+\nu^2x^2}{(1-\nu^2x^2)^2}}$
        \\\hline
        \end{tabular}
    \tablefoot{$r_{\zeta x}$ and $r_{x x}$ denote  $\partial^2_{\zeta x} r$ and $\partial^2_{x x} r$, respectively and $f$ is a function of $x$ given by the stellar model.}
\end{table*}

In our work, we solve for each spheroid only one linear ordinary differential equation (the generalised Laplace tidal equation) which depends on the colatitude and in a parametric way on the pseudo-radius. Indeed, since we focus on low-frequency gravito-inertial waves, which are rapidly oscillating along the pseudo-radial direction, we adopt the JWKB approximation. It allows us to describe the spatial structure of the waves by the product of a rapidly oscillating plane-like wave function in the pseudo-radial direction multiplied by a bi-dimensional slowly varying envelope (Eqs.\;\ref{eq:jwkb_w} \& \ref{eq:jwkb_u}) which is obtained by solving the GLTE (Eq.\;\ref{eq:glte}) for each spheroidal shell for the normalised pressure and polarisation relationships for the different components of the velocity (Eqs.\;\ref{eq:u_zeta_final} - \ref{eq:u_phi_final}). Therefore, we avoid having an ordinary differential equation to solve in the pseudo-radial direction. This method has been introduced and used by \cite{mathis2009} and \cite{vanreeth2018} in the case of the generalisation of the TAR to the case of spherical differentially rotating stars and by \cite{mathis+prat2019} and \cite{Henneco2021} in the case of the generalisation of the TAR for slightly deformed rotating stars. Finally, solving the GLTE allows us to compute the horizontal eigenvalues and the corresponding eigenfrequencies.

\subsection{Asymptotic frequency and period spacing of low-frequency GIWs}
From Eqs.\;(\ref{eq:continuity_final}) and (\ref{eq:glte}), we can identify the dispersion relation for low-frequency GIWs within the TAR for strongly deformed stars
\begin{equation}\label{eq:dispersion}
    k_{V ; \nu k m}^{2}(\zeta)=\frac{N^2(\zeta)\mathcal{A}(\zeta)}{\omega_{k m}^{2}} \Lambda_{\nu k m}(\zeta).
\end{equation}
To be able to derive the eigenfrequencies of low-frequency GIWs, we applied the quantisation relation in the vertical (pseudo-radial) direction used by \citet{mathis2009} and \citet{mathis+prat2019} and introduced by \cite{unno1989}, \cite{gough1993}, and \cite{Christensen1997}

\begin{equation}\label{eq:quantisation}
    \int_{\zeta_{1}}^{\zeta_{2}} k_{V ; \nu n k m} \mathrm{d} \zeta=(n+1 / 2) \pi,
\end{equation}
where $\zeta_1$ and $\zeta_2$ are the turning points of the 
Brunt–Väisälä frequency $N$ and $n$ is the radial order.

So, by substituting the dispersion relation (Eq.\;\ref{eq:dispersion}) into the vertical quantisation one (Eq.\;\ref{eq:quantisation}), we get the asymptotic expression for the frequencies of low-frequency GIWs
\begin{equation}\label{eq:frequencies}
    \omega_{n k m}=\frac{\int_{\zeta_{1}}^{\zeta_{2}} N(\zeta)\sqrt{\mathcal{A}(\zeta)\Lambda_{\nu_{n} k m}(\zeta)} \mathrm{d} \zeta}{(n+1 / 2) \pi},
\end{equation}
and the corresponding period
\begin{equation}\label{eq:period}
    P_{n k m}=\frac{2(n+1 / 2) \pi^2}{\int_{\zeta_{1}}^{\zeta_{2}} N(\zeta)\sqrt{\mathcal{A}(\zeta)\Lambda_{\nu_{n} k m}(\zeta)} \mathrm{d} \zeta},
\end{equation}
with 
\begin{equation}
    \nu_n=\nu_{nkm}=\frac{2 \Omega}{\omega_{nkm}}.
\end{equation}
We can thus compute the period spacing $\Delta P _{km}= P_{n+1km}- P_{nkm}$. We note that by applying a first-order Taylor development we can generalise the result obtained by \cite{bouabid2013} in the spherical case
\begin{equation}\label{eq:period_spacing}
    \Delta P _{km}\approx\frac{2 \pi^2}{\int_{\zeta_{1}}^{\zeta_{2}} N\sqrt{\mathcal{A}\Lambda_{\nu_{n+1} k m}} \mathrm{d} \zeta \left(1+\frac{1}{2}\frac{\int_{\zeta_{1}}^{\zeta_{2}} N\sqrt{\mathcal{A}\Lambda_{\nu_{n} k m}} \frac{d \ln \Lambda_{\nu_{n} k m}}{d \ln{\nu}} \mathrm{d} \zeta}{\int_{\zeta_{1}}^{\zeta_{2}} N\sqrt{\mathcal{A}\Lambda_{\nu_{n} k m}} \mathrm{d} \zeta}\right)},
\end{equation}
which will potentially allow us to probe the internal rotation of a large number of rapidly rotating stars taking into account their possibly strong flattening by the centrifugal acceleration and perform their detailed seismic modelling.

This result generalises and is very similar to the one obtained in the spherical case \citep{lee+saio1997, bouabid2013} and in the weakly deformed case using a perturbative approach \citep{mathis+prat2019}. The major interest of the eigenfrequencies, periods and periods spacing (Eqs.\;\ref{eq:frequencies}-\ref{eq:period_spacing}) is that they can be easily evaluated for a large number of rapidly rotating stars and compared to observations which is a great asset to perform detailed seismic modelling. This flexibility is  what makes the TAR a very interesting treatment.

\section{Application to rapidly rotating early-type stars}\label{sect:results}
In order to implement our equations and solve the GLTE, we must first compute an equilibrium model which gives the structure of the rotating star. The most advanced model of stellar evolution that we have and which is adapted to the formalism that we developed here is the ESTER model \citep{espinosa+rieutord2013}. It is unique of its kind because it calculates the effect of the centrifugal acceleration in a non-perturbative and a self-consistent way at any rotation rate.

\subsection{ESTER models}
The ESTER code is a 2D stellar structure code which takes into account the effects of rotation in a non-perturbative and coherent manner.
The ESTER code makes it possible to calculate the equilibrium state of a star with a mass greater than 2 $M_{\odot}$ by solving the stationary equations of hydrodynamics for pressure, density, temperature and angular velocity for the entire volume. These realistic two-dimensional models of rapidly rotating stars are computed with the actual baroclinic meridional flows and differential rotation for intermediate-mass and massive stars. In this work, as a first step we work only with a uniform rotation (the mean value of the differential rotation over the pseudo-radius and the colatitude). The inclusion of differential rotation in the TAR formalism will be done in Paper II.
The code uses spectral methods in the radial and latitudinal directions to transform the original system of non-linear partial differential equations in a system of non-linear algebraic equations (Chebyshev polynomials in the pseudo-radial direction and  with spherical harmonics in the horizontal direction) \citep{rieutord+espinosa2013}.

In this study, we use $3 \mathrm{M}_{\odot}$ stellar models with a hydrogen mass fraction in the core $X_{\rm c} = 0.7$ at $[0\%,90\%]$ of the Keplerian break-up rotation rate $\Omega_{\mathrm{K}}$. Such a model is representative of observed stars near the zero-age main-sequence with asteroseismic probes of their angular velocity \citep{vanreeth2015,vanreeth2016,vanreeth2018,Papics2015,Papics2017,li2019a,li2019b}.

\subsection{Domain of validity of the TAR} \label{subsect:validity}
In this section, we study the validity domain of the two approximations  (\ref{subsect:hypothesis1} and \ref{subsubsect:hypothesis2}) that are necessary to build the TAR in the case of rapidly rotating deformed bodies:
\begin{gather}
    N^2(\zeta, \theta)\approx N^2  (\zeta),\label{eq:approx1}\\
    \mathcal{A}(\zeta, \theta)\approx\mathcal{A}(\zeta).\label{eq:approx2}
\end{gather}

\begin{figure}
    \centering
    \resizebox{\hsize}{!}{\includegraphics{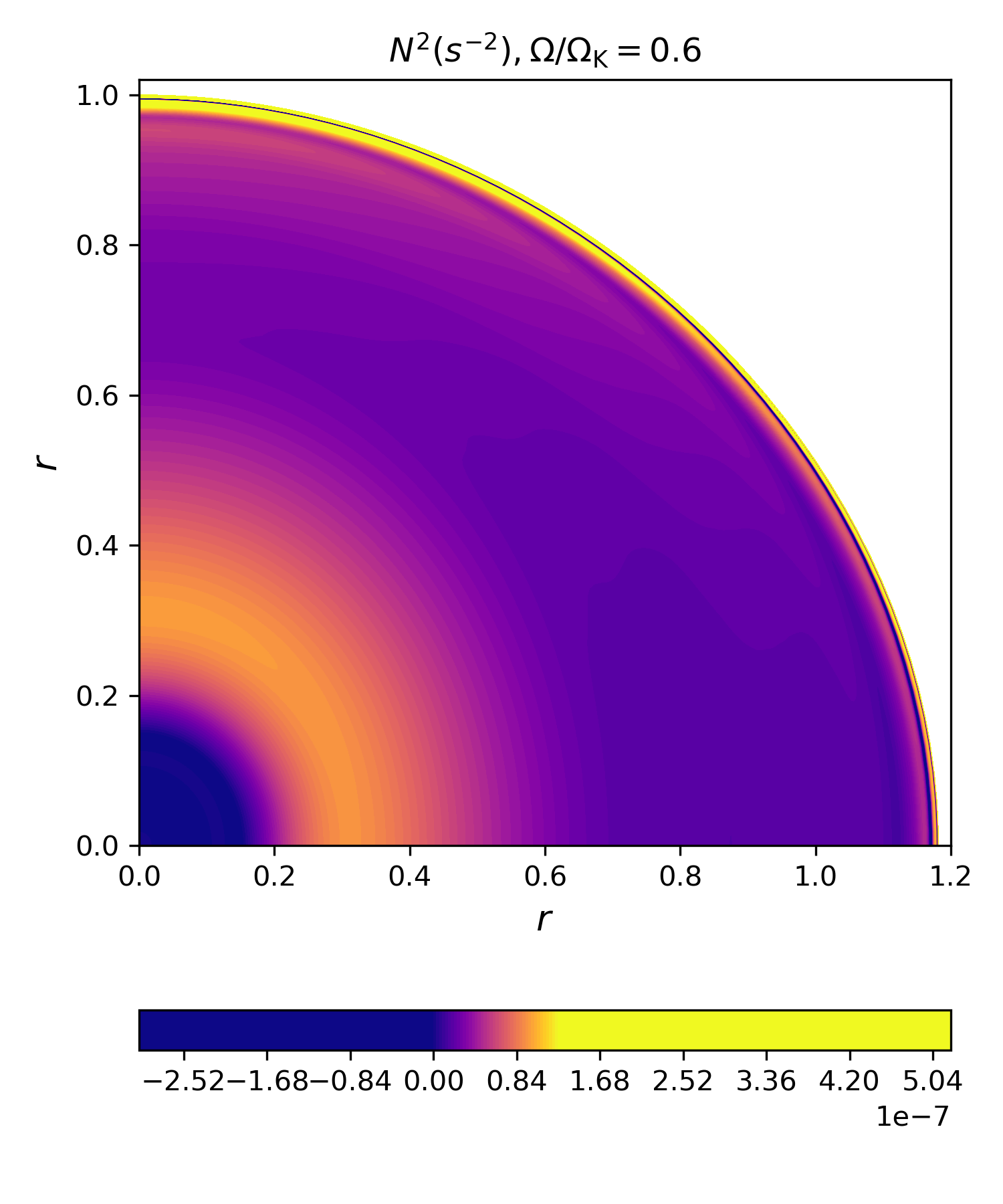}}
    \caption{Squared Brunt–Väisälä frequency  $N^2(\zeta,\theta)$ profile for $\Omega/\Omega_{\rm K}=60\%$.}
    \label{fig:N2_2D}
\end{figure}

\begin{figure}
    \centering
    \resizebox{\hsize}{!}{\includegraphics{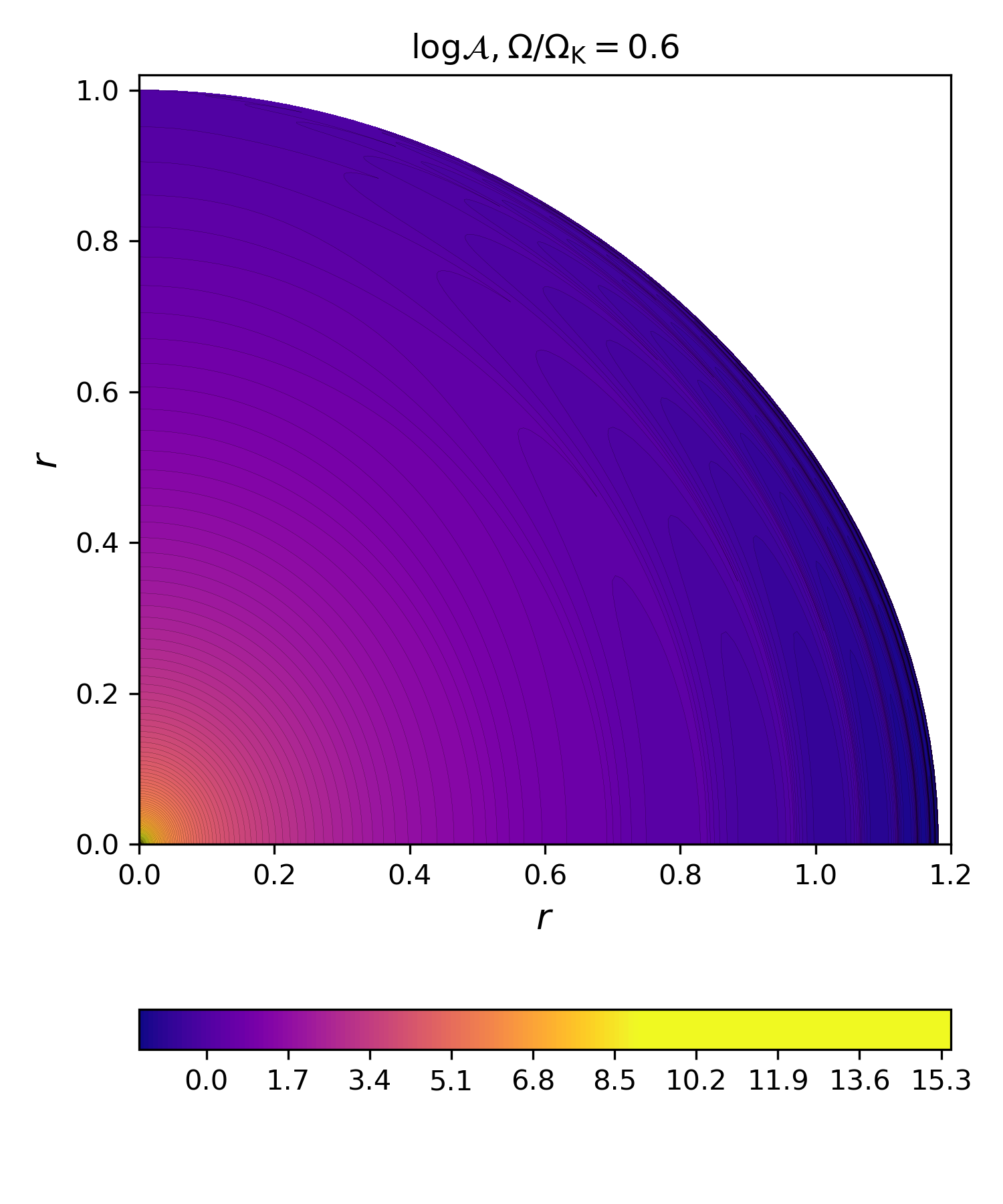}}
    \caption{Coefficient $\mathcal{A}(\zeta,\theta)$ profile for $\Omega/\Omega_{\rm K}=60\%$.}
    \label{fig:A_2D}
\end{figure}

To visualise the problem, we first illustrate in Figs.\;\ref{fig:N2_2D}\;\&\;\ref{fig:A_2D} the functions $N^2(\zeta, \theta)$ and $\mathcal{A}(\zeta, \theta)$ computed with an ESTER model rotating at $\Omega/\Omega_{\rm K}=60\%$ (we use $r$ instead of $\zeta$ in these figures so we can show the spheroidal shape of the star caused by the centrifugal deformation). We can see from Fig.\;\ref{fig:N2_2D} that the convective region of the star is located between $\zeta=0$ and $\zeta=0.153$ and that the pseudo-radial variation of the Brunt–Väisälä frequency is the highest at the surface. From Fig.\;\ref{fig:A_2D} we can see that the coefficient $\mathcal{A}$ varies the most, according to the pseudo-radius, at the centre of the star where we approach its singularity. In order to be able to apply our approximations  (Eqs.\;(\ref{eq:approx1})\;\&\;(\ref{eq:approx2})), we are interested in quantifying the latitudinal variation of this two quantities.

\begin{figure}
    \centering
    \resizebox{\hsize}{!}{\includegraphics{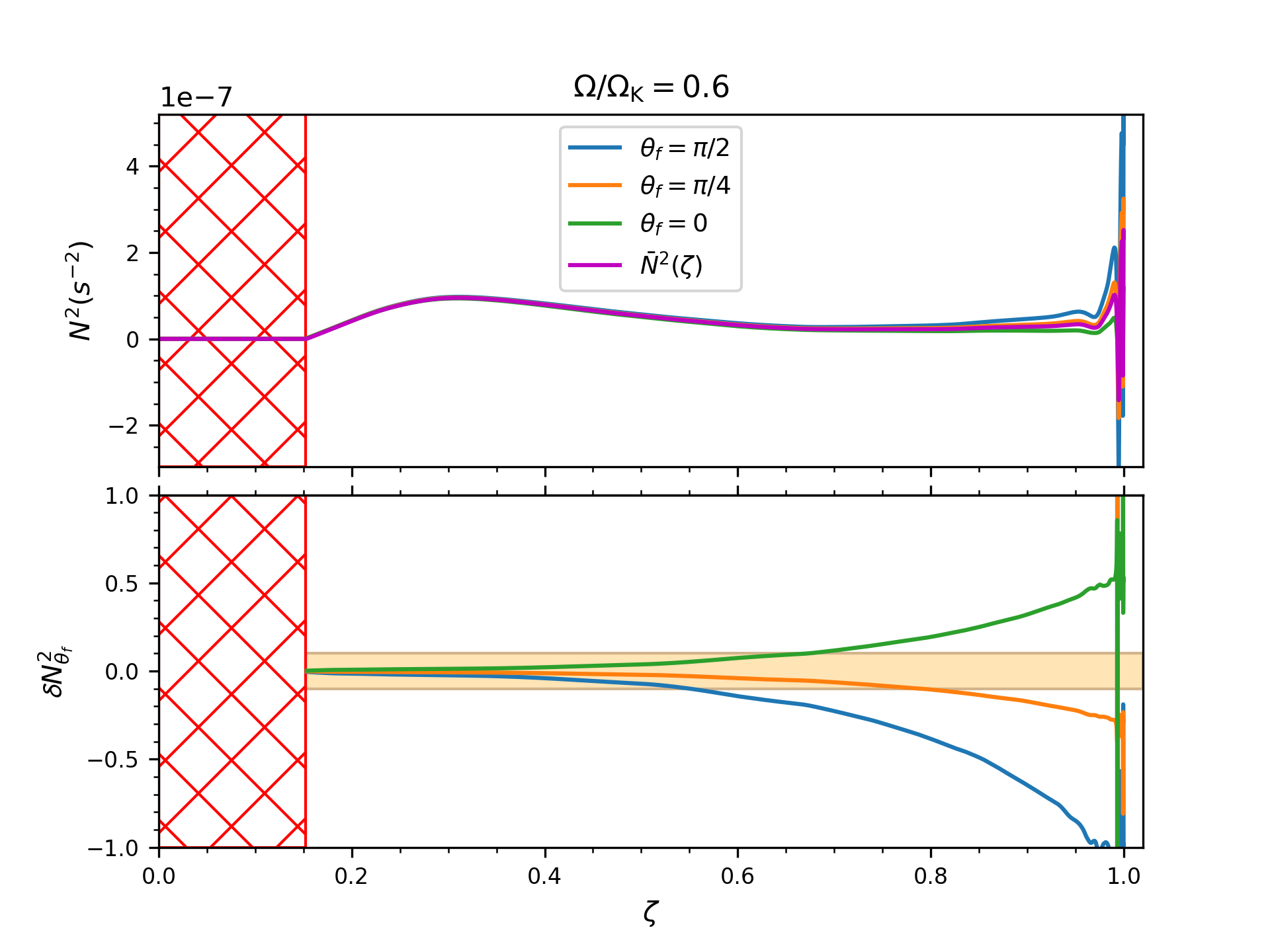}}
    \caption{Profile of the squared Brunt–Väisälä frequency $N^2$ and the relative error $\delta N^2_{\theta_f}$ of the adopted approximation (Eq.\;\ref{eq:approx1}) as a function of $\zeta$ at different colatitudes $\theta_f$ using an ESTER model rotating at $60\%$ of the Keplerian breakup rotation rate (The light orange area indicates the margin of error which we allow and the red hatched area represents the convective region of the star).}
    \label{fig:N-0.6}
\end{figure}

\begin{figure}
    \centering
    \resizebox{\hsize}{!}{\includegraphics{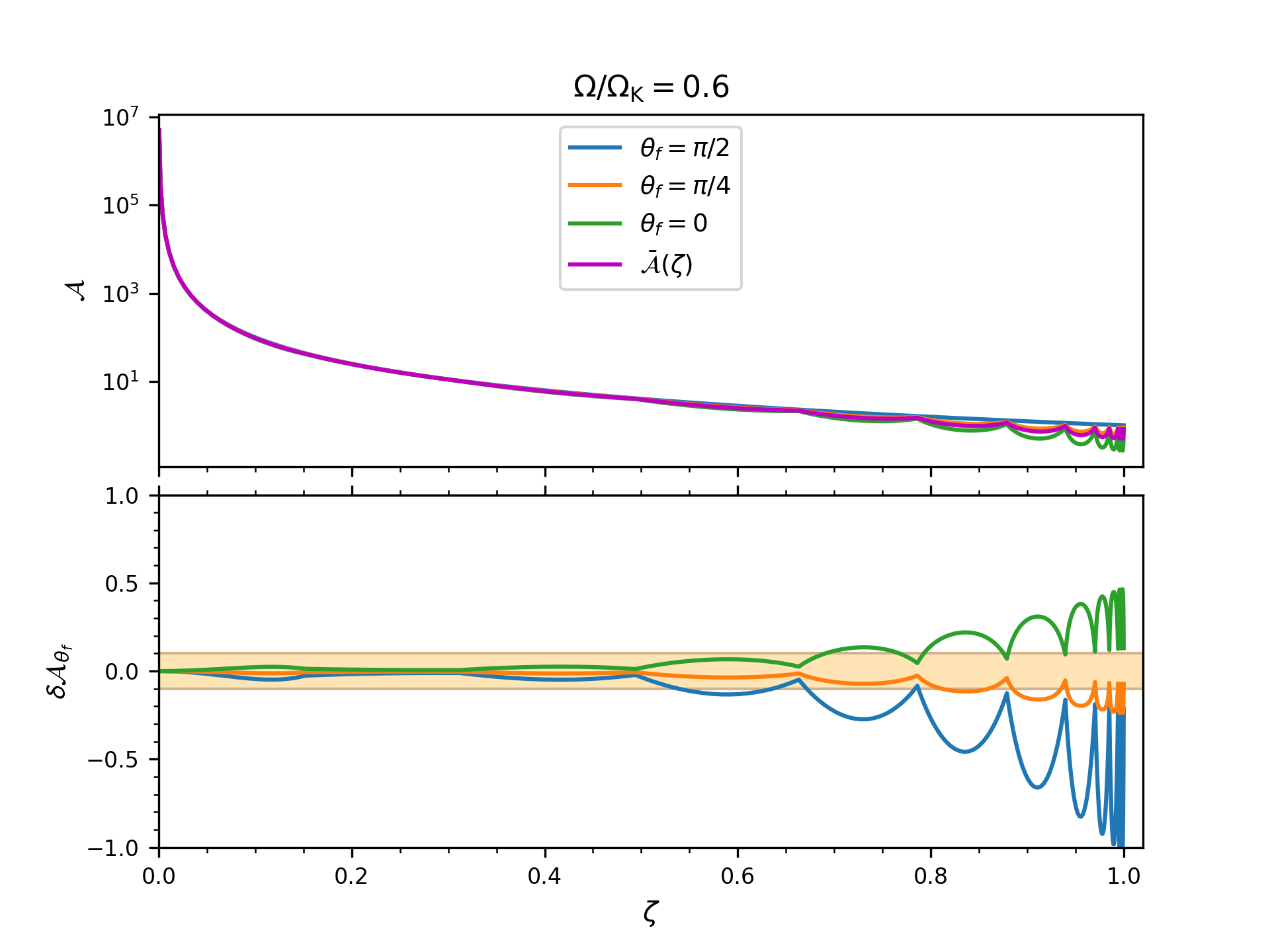}}
    \caption{Profile of the coefficient $\mathcal{A}$ and the relative error $\delta \mathcal{A}_{\theta_f}$ of the adopted approximation (Eq.\;\ref{eq:approx2}) as a function of $\zeta$ at different colatitudes $\theta_f$ using an ESTER model rotating at $60\%$ of the Keplerian break-up rotation rate (The light orange area indicates the margin of error which we allow).}
    \label{fig:A-0.6}
\end{figure}

So to evaluate the committed error by adopting Eqs.\;(\ref{eq:approx1})\;\&\;(\ref{eq:approx2}) we represent in Figs.\;\ref{fig:N-0.6}\;\&\;\ref{fig:A-0.6} the 
Brunt–Väisälä frequency and the coefficient $\mathcal{A}$ pseudo-radial profiles for different values of the colatitude $\theta_f$, at a fixed rotation rate $\Omega=0.6 \Omega_{\rm K}$. In the bottom panels, we show the  corresponding relative error 
\begin{equation}
    \delta X_{\theta_f} (\zeta)= \frac{X_{\rm approx}(\zeta) - X(\zeta,\theta_f)}{X(\zeta,\theta_f)},
\end{equation}
between the exact value $X(\zeta,\theta_f)$ given by the model and the approximated model $X_{\rm approx}(\zeta)$ given by the weighted average of the exact value  over the colatitude $\theta$
\begin{equation}
    X_{\rm approx}(\zeta)=\bar{X}(\zeta)=\int_0^{\pi/2}X(\zeta,\theta)\sin\theta d\theta,
\end{equation}
with
\begin{equation*}
    X \equiv\{N^2, \mathcal{A}\},
\end{equation*}
where $\theta_f$ is a fixed value of the colatitude $\theta$.


\begin{figure}
    \centering
    \resizebox{\hsize}{!}{\includegraphics{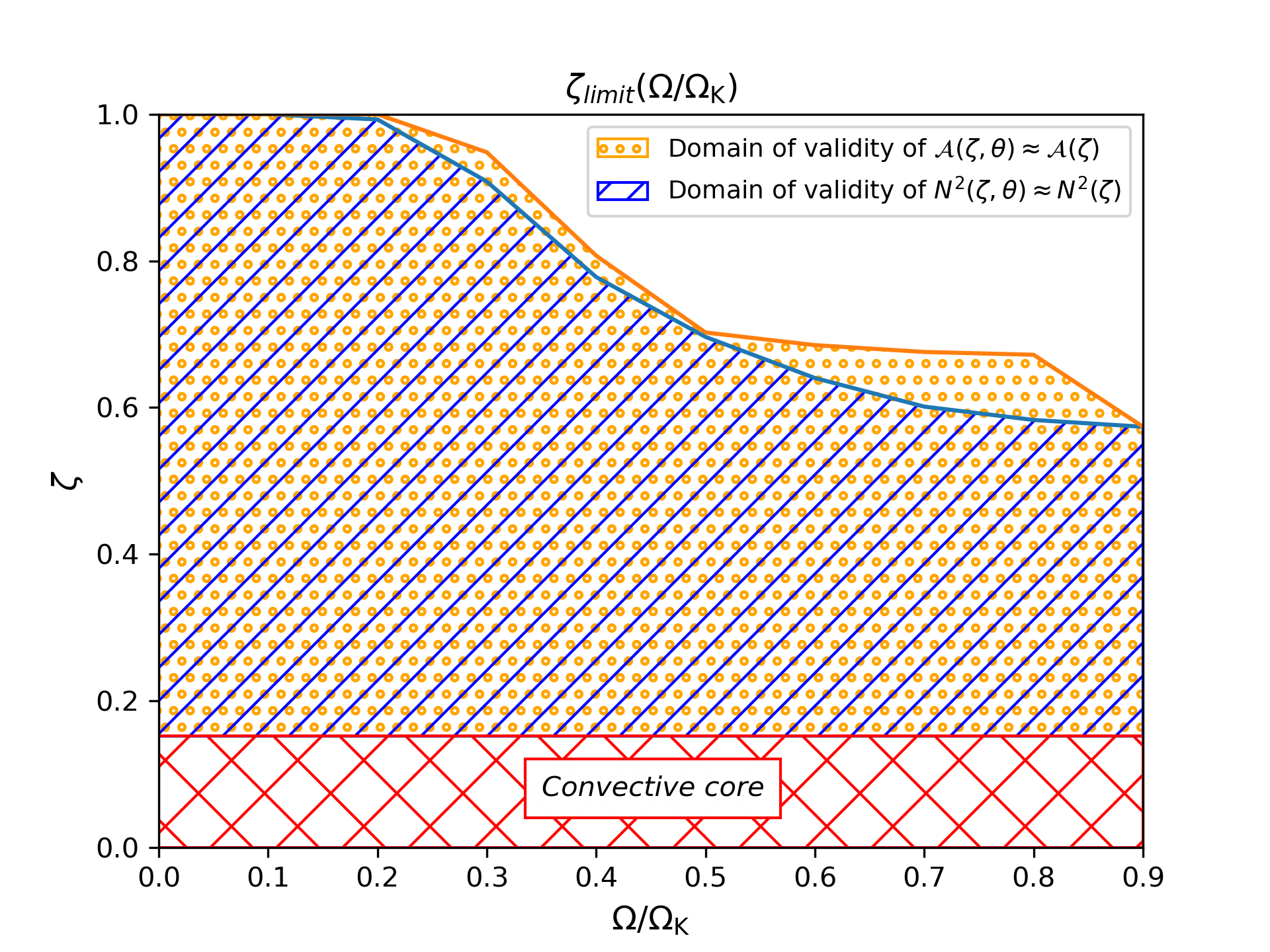}}
    \caption{Validity domain of the TAR within the framework of ESTER models ($3\,\mathrm{M}_{\odot}$, $X_{\mathrm{c}}=0.7$) as a function of the rotation rate $\Omega /\Omega_{\rm K}$ and the pseudo-radius $\zeta$ (with $90\%$ degree of confidence).}
    \label{fig:domain}
\end{figure}

Then, in order to determine the validity domain of these approximations as a function of the rotation rate and the pseudo-radius, we vary the rotation rate of the star $\Omega/\Omega_{\rm K}$ (we use here 10 sample points for the rotation rate) and we calculate for each case the associate maximum relative error $\max_{i}\delta X_{\theta_i} (\zeta)$. Afterwards, we fix a maximum error rate equal to $10\%$ and we deduce the pseudo-radius limit $ \zeta_{\rm limit} $ where the maximum relative error exceeds this threshold and we decide that the approximations become invalid. In Fig.\;\ref{fig:domain}, we display the pseudo-radius limit $ \zeta_{\rm limit} $ as a function of the rotation rate $\Omega/\Omega_{\rm K}$. This curve defines the validity domains of the two approximations (Eqs.\;\ref{eq:approx1}\;\&\;\ref{eq:approx2}) which consequently define the validity domain of the TAR.

We notice that the validity domains of the two approximations are more and more restricted (the pseudo-radius limit $\zeta _{\rm limit} $ decrease) by increasing the rate of rotation and approaching the critical angular velocity. The tendency of the reduction of the pseudo-radius limit with the increasing of the rotation rate is expected since the action of the centrifugal acceleration is stronger further away from the axis of rotation. Thus, to be able to apply the TAR in the whole space domain, we shouldn't exceed a rotation rate $\Omega = 0.2\Omega_{\rm K}$. It is therefore possible to apply the generalised TAR to early-type stars ($3\,\mathrm{M}_{\odot}$, $X_{\mathrm{c}}=0.7$) with an external radiative zone rotating up to 20\% of the Keplerian critical angular velocity. This limit is lower than the one proposed by \citet{mathis+prat2019} which was $\Omega=0.4\Omega_{\rm K} $. The limit proposed here should be more realistic since it is derived from a two-dimensional model which takes into account the centrifugal acceleration in a non-perturbative way whereas \citet{mathis+prat2019} treated it in a perturbative manner. 
In the case where $\zeta_{\rm limit}<1$, generalised TAR should also be applicable to the radiative core of rapidly rotating (young) solar-type stars for which $\zeta_{\rm rad}<\zeta_{\rm limit}$, where $\zeta_{\rm rad}$ is the pseudo radius of the radiation-convection interface. So, this formalism can be applied to all rapidly rotating bodies (stars of all types and planets) as long as we have a 2D model with access to the  quantities that we used to derive our equations.

\subsection{Eigenvalues and Hough functions}
We solve the GLTE for different pseudo-radii, spin parameters, and rotation rates within the defined validity domain using the implementation based on Chebyshev polynomials presented and discussed by \citet{wang2016} and then used by \citet{mathis+prat2019} and \cite{Henneco2021}.

\begin{figure}
    \centering
     \resizebox{\hsize}{!}{\includegraphics{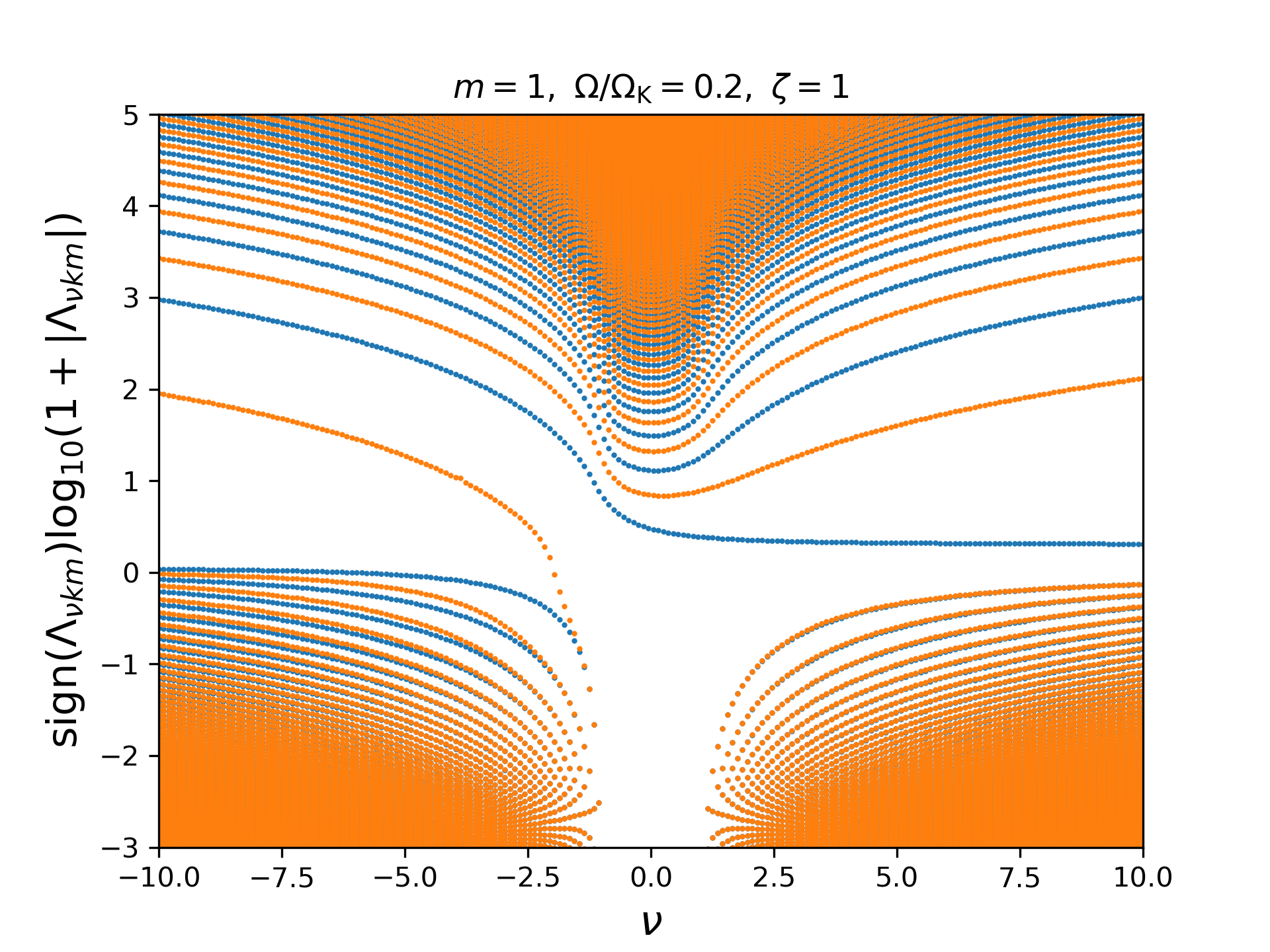}}
     \resizebox{\hsize}{!}{\includegraphics{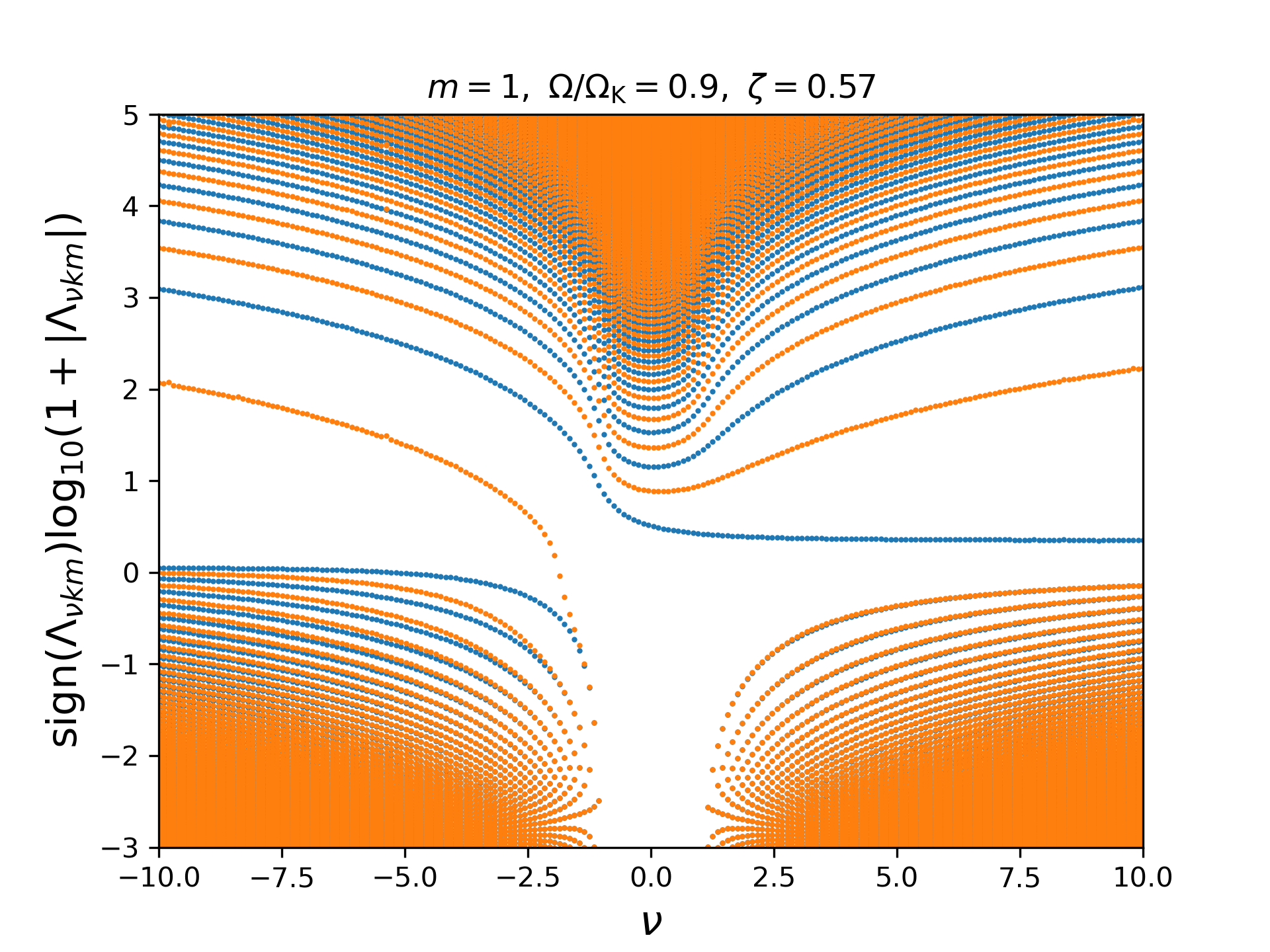}}
    \caption{Spectrum of the GLTE as a function of the spin parameter $\nu$ at $\zeta=\zeta_{\rm limit}$ and $m=1$ for $\Omega=0.2 \Omega_{\mathrm{K}}$ (above) and $\Omega=0.9 \Omega_{\mathrm{K}}$ (below).  Blue (respectively, orange) dots correspond to even (respectively, odd) eigenfunctions.}
    \label{fig:spectrum_z1_o0.2}
\end{figure}

Fig.\;\ref{fig:spectrum_z1_o0.2} shows the eigenvalues $ \Lambda_{\nu km}$ of the GLTE at $\zeta=\zeta_{\rm limit}$ as a function of the spin parameter $\nu$ for $m = 1$ at two different rotation rates: $\Omega = 0.2\Omega_{\rm K}$ the maximum rotation rate for which the generalised TAR is valid in all the radiative region ($\zeta_{\rm limit}=1$) and $\Omega = 0.9\Omega_{\rm K}$ the highest rotation rate that we consider in this study for which the generalised TAR is valid in a restricted space domain ($\zeta_{\rm limit}=0.57$). Since $\vec{\Omega}$ points in the direction of $\theta =0$ and the oscillations are proportional to $e ^{i (\omega t - m \varphi)}$, a prograde (retrograde) oscillation  has a positive (negative) value of the product $m\nu$. Therefore, since $m =1 $ the prograde and retrograde modes correspond respectively to the positive and negative values of $\nu$. We see here that the centrifugal deformation of the star causes a modest gradual shift in the eigenvalues as already identified in previous studies where the centrifugal acceleration has been treated in a perturbative way \citep{mathis+prat2019,Henneco2021}.

\begin{figure}
    \centering
     \resizebox{\hsize}{!}{\includegraphics{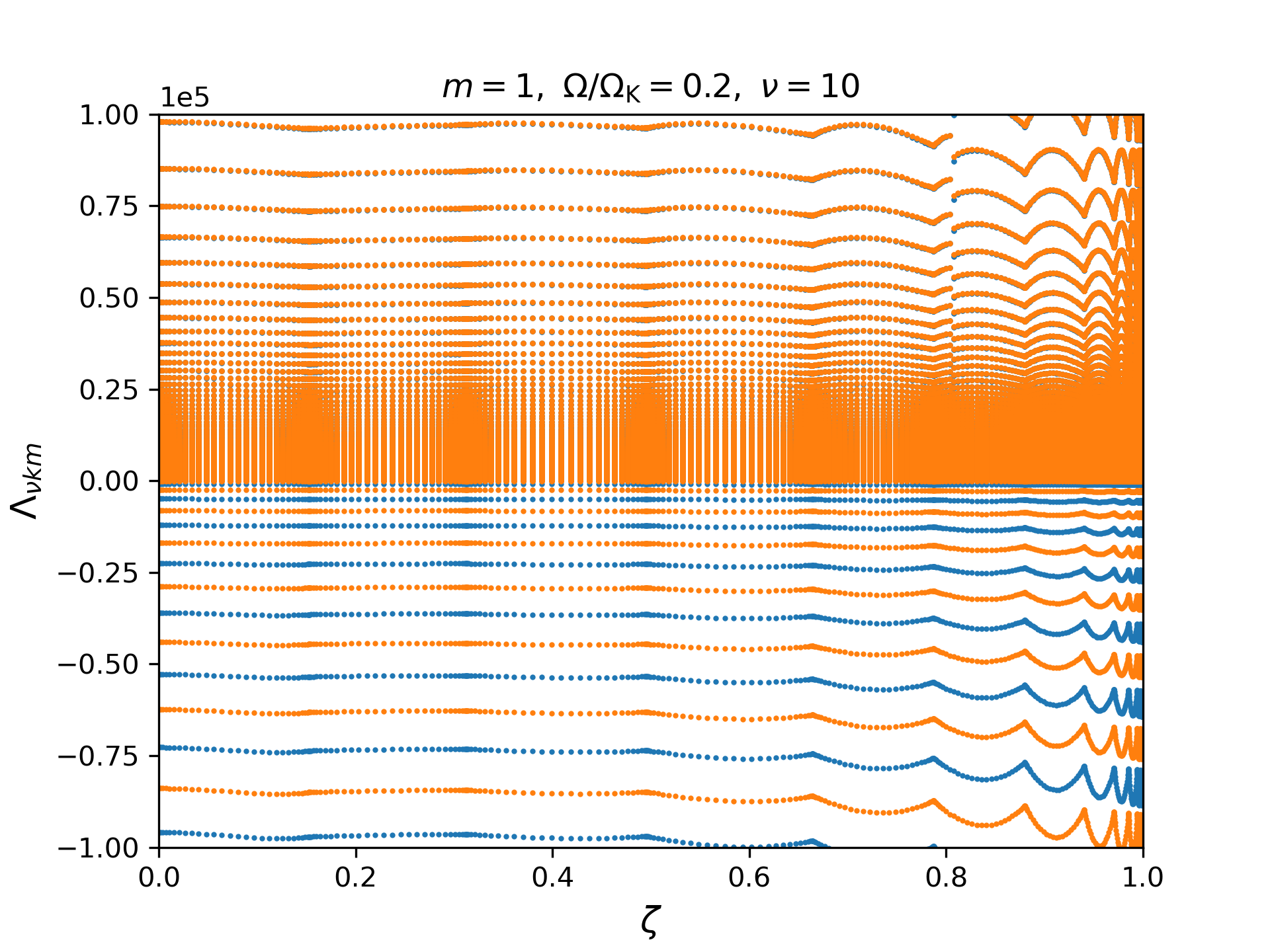}}
     \resizebox{\hsize}{!}{\includegraphics{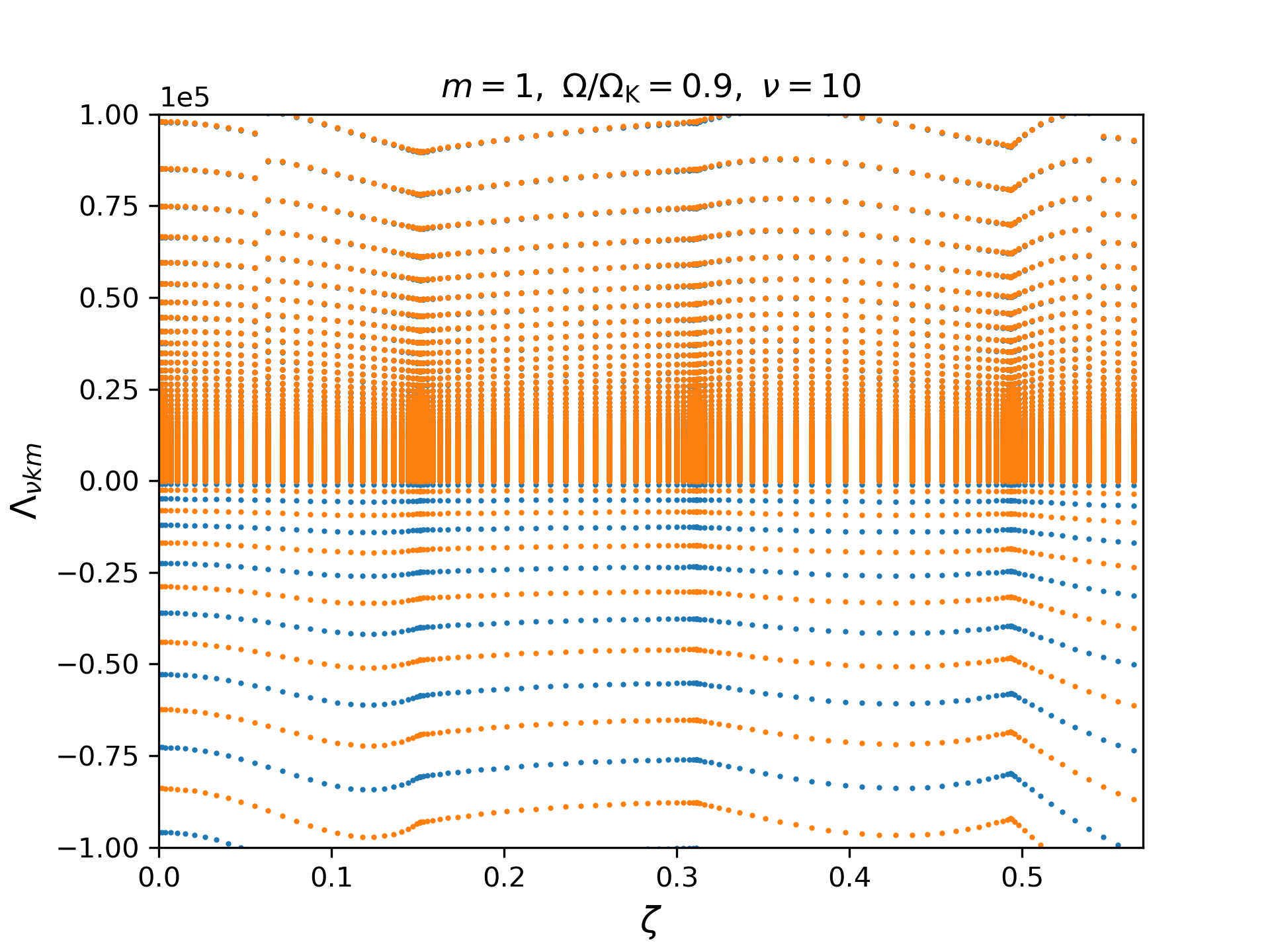}}
    \caption{Spectrum of the GLTE as a function of the pseudo-radius for $m=1$ and $\nu=10$ at $ \Omega/\Omega_K = 0.2 $ (above) and $ \Omega/ \Omega_K= 0.9 $ (below). Blue (respectively, orange) dots correspond to even (respectively, odd) Hough functions.}
    \label{fig:spectrumz_nu5}
\end{figure}

We can distinguish two families of eigenvalues. 
Namely, the first one is gravity-like solutions ($\Lambda_{\nu k m}\geqslant0$) which exist for any value of $\nu$ and we attach positive $k$'s to them. They correspond to gravity waves ($\mathrm{g}$ modes) modified by rotation. The second one is Rossby-like solutions which exist only when $|\nu|> 1$ and we attach negative $k$'s to them. They only propagate in rotating stars. They correspond to Rossby modes ($\mathrm{r}$ modes) if they are retrograde and have positive eigenvalues ($\Lambda_{\nu k m}>0$) \citep{saio2018} and to  overstable convective modes if they are prograde and have negative eigenvalues ($\Lambda_{\nu k m}<0$). These modes are able to propagate in convective regions stabilised by the rotation under the action of the Coriolis acceleration \citep{lee+saio1997,lee2019,lee+saio2020}. Even though the TAR is not a valid approximation in convective zones \citep{ogilvie+lin2004}, we do find such propagative modes as solutions from our approach.
Contrary to the spherical case we obtain here eigenvalues which depend slightly on the pseudo-radius $\zeta$ as shown in Fig.\;\ref{fig:spectrumz_nu5} where we represent the spectrum of the GLTE as a function of the pseudo-radius at two different rotation rates  $ \Omega/\Omega_K =\{20\%, 90\%\}$ for $m=1$ and $\nu=10$.

\begin{figure}
    \centering
     \resizebox{\hsize}{!}{\includegraphics{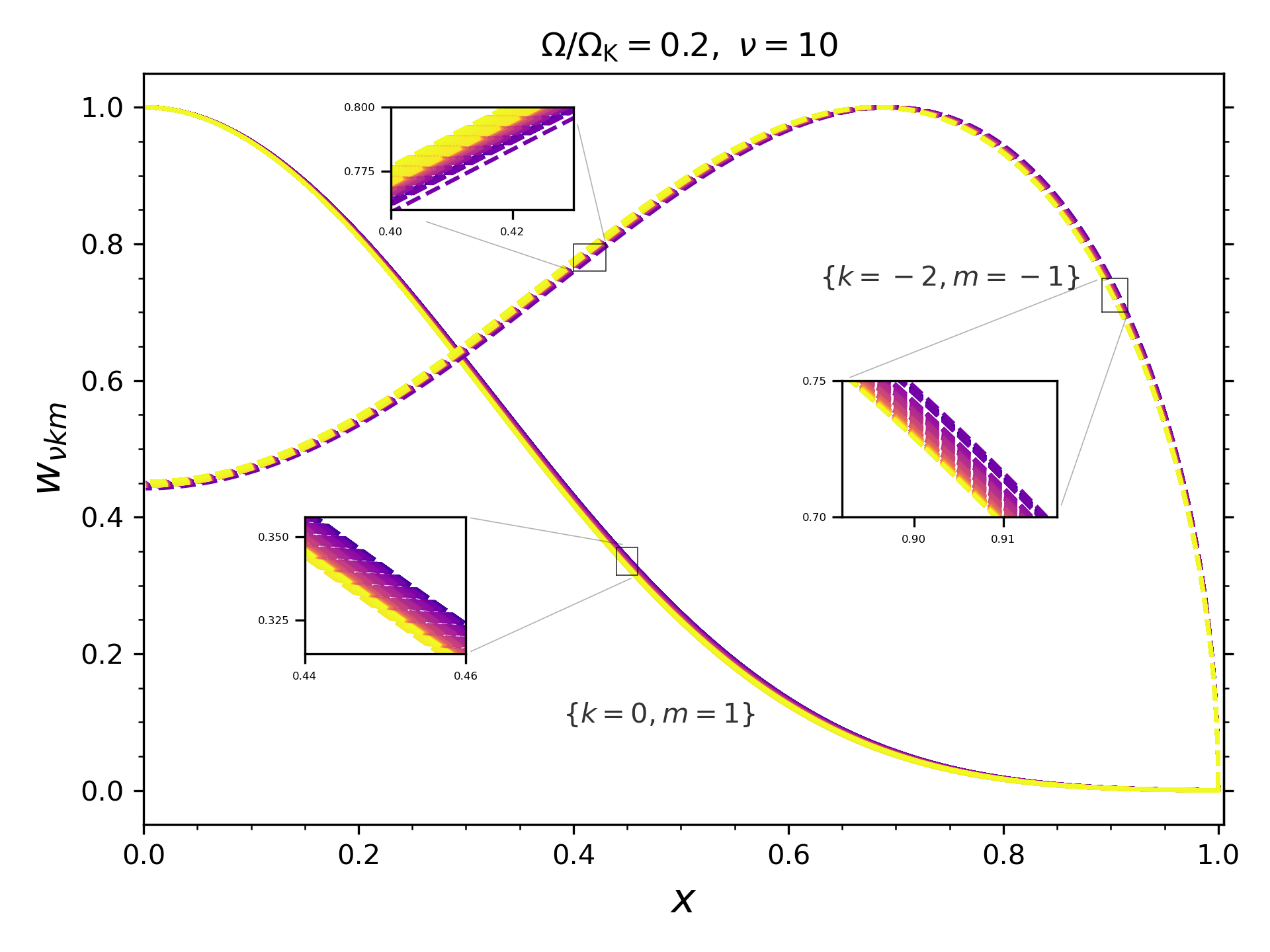}}
     \resizebox{\hsize}{!}{\includegraphics{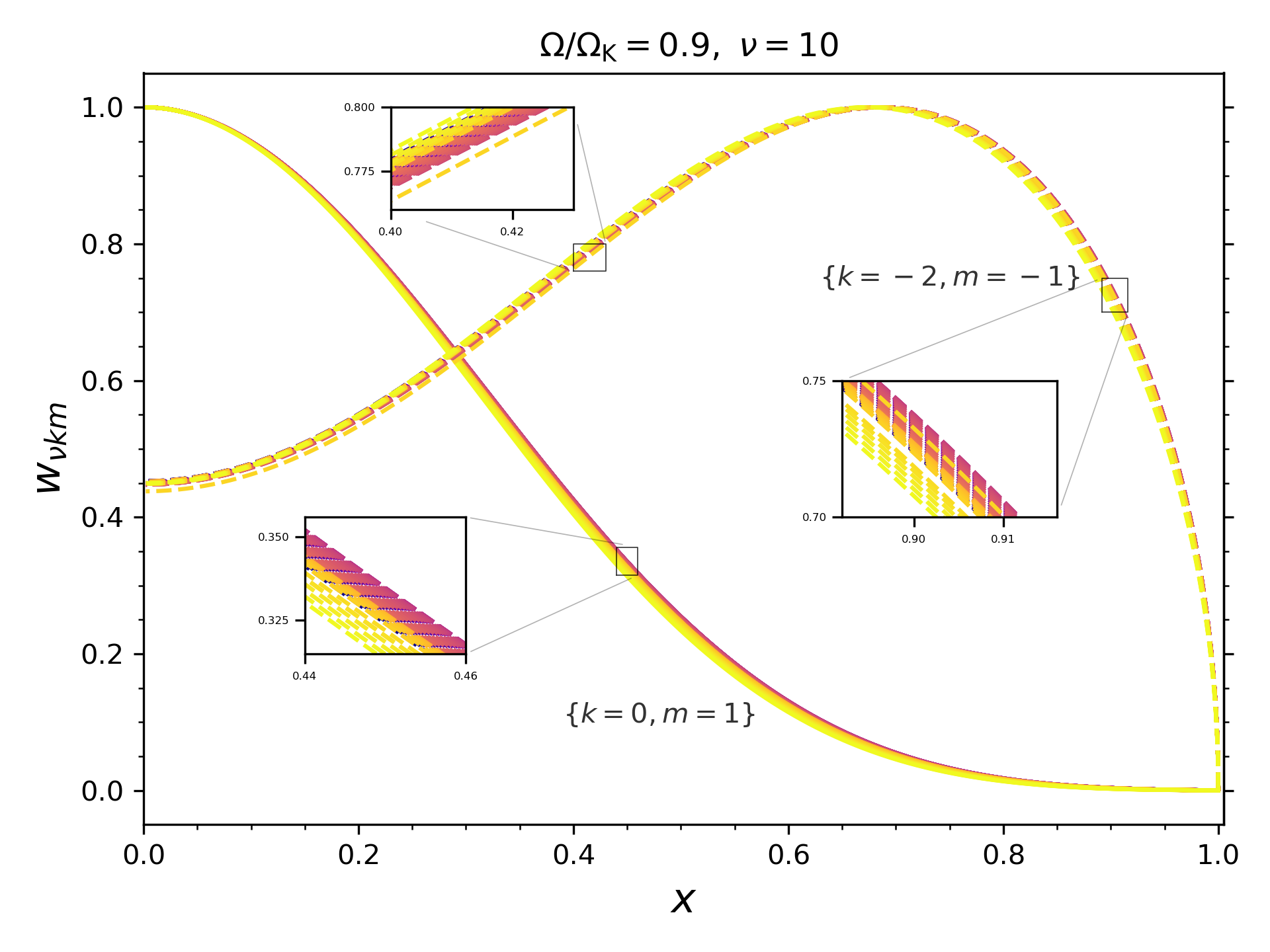}}
    \caption{Generalised Hough functions (normalised to unity) as a function of the horizontal coordinate $x$ at different pseudo-radii from $\zeta=0.153$ (blue) to $\zeta=\zeta_{\rm limit} $ (yellow) for $\nu=10$, $ \Omega/\Omega_K = 0.2 $ (above), and $ \Omega/ \Omega_K= 0.9 $ (below). The  solid lines correspond to gravity-like solutions with $ \{k = 1, m=1\} $, while the dotted lines correspond to Rossby-like solutions with $ \{k = -2, m=-1\} $.}
    \label{fig:hough_nu5}
\end{figure}

We focus now on the  generalised Hough functions $w_{\nu km}$ (normalised to unity) which vary with the pseudo-radius $\zeta$ and the horizontal coordinate $x$. This dependence is illustrated at $\nu= 10$ in Fig.\;\ref{fig:hough_nu5} at two different rotation rates $ \Omega/\Omega_K =\{20\%, 90\%\}$ for gravity like-solutions (prograde dipole $\{k = 0, m=1\}$ modes) and Rossby like-solutions (retrograde  Rossby  modes  with $\{k = -2, m=-1\}$). 
We can see that increasing the rotation rate from 20\% to 90\% of the critical angular velocity does hardly affect the Hough functions. This is caused by the fact that increasing the rotation rate implies decreasing the pseudo-radius limit ($\zeta_{\rm limit}$) so we can no longer trust the computed Hough functions near the surface where the centrifugal effect is maximal.
We can see clearly in this figure that the centrifugal acceleration slightly influences  and modifies the eigenfunctions of the GLTE by introducing a new dependency on the pseudo-radial coordinate $\zeta$, whereas in the spherically symmetric case they were only dependant on the latitudinal coordinate $\theta$. In fact, the gravity-like solutions and the Rossby-like solutions shift as the distance from the border of the radiative zone of the star ($\zeta=0.153$) to its surface increases ($\zeta=1$). More specifically, we observe a non-monotonous behaviour in Rossby-like solutions: these solutions migrate at first onwards, away from the equator ($x = 0$), causing a broadening of its shape. Next, by increasing the reduced latitudinal coordinate $x$ this behaviour becomes less and less visible until it changes to the opposite one where these solutions migrate inwards, towards the equator, causing a narrowing of its shape. However, in gravity-like solutions we found a monotonous behaviour for which these solutions migrate inwards, towards the equator, causing a narrowing of its shape. Contrary to the non-monotonous behaviour found in Rossby-like solutions, \cite{Henneco2021} found using the perturbative TAR a monotonous one with a migration towards the equator in the two types of solutions.

\section{Asymptotic seismic diagnosis}\label{sect:seismic_diagnosis}
\subsection{Asymptotic period spacing pattern}
\begin{figure}
    \centering
    \resizebox{\hsize}{!}{\includegraphics{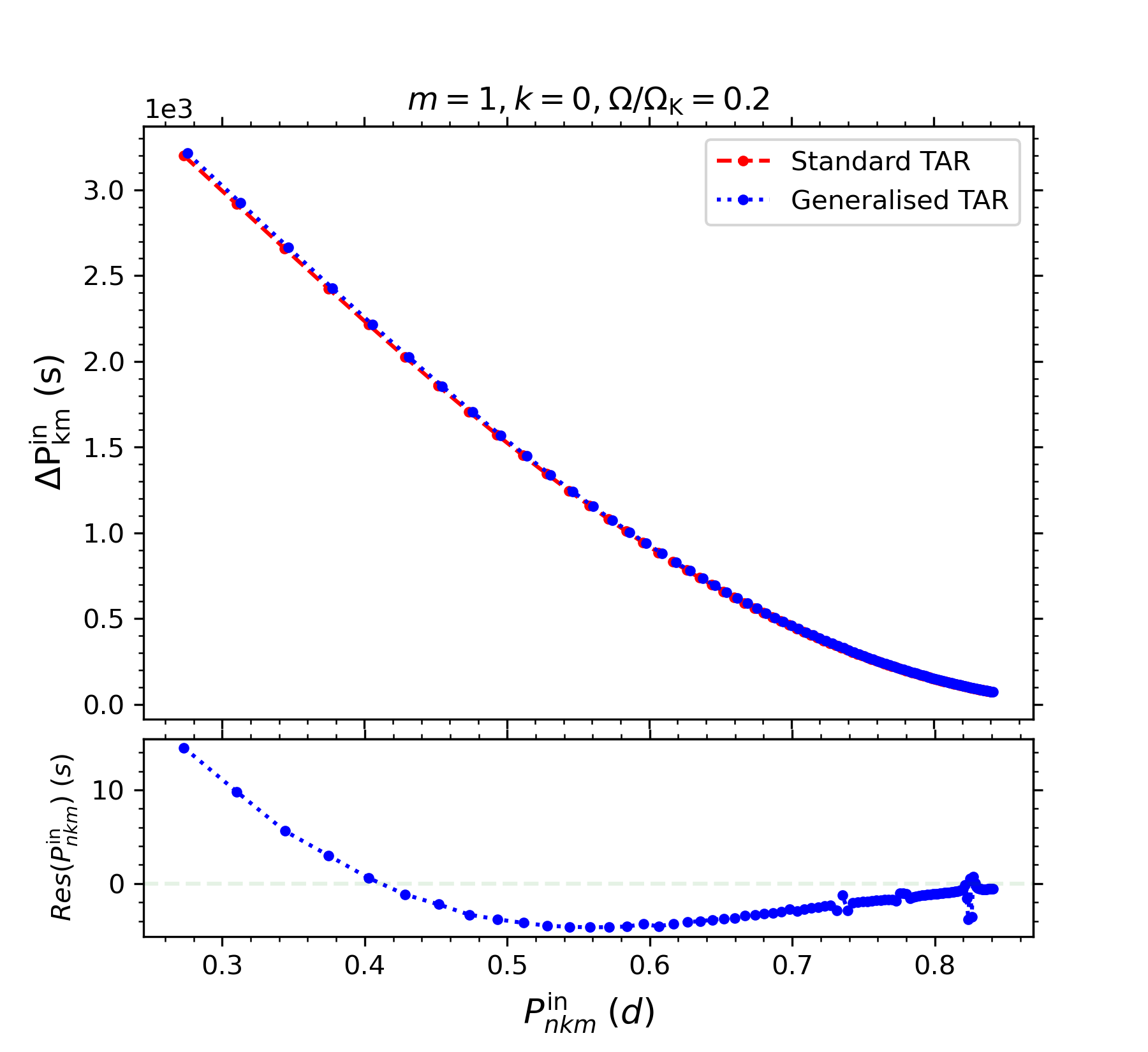}}
    \resizebox{\hsize}{!}{\includegraphics{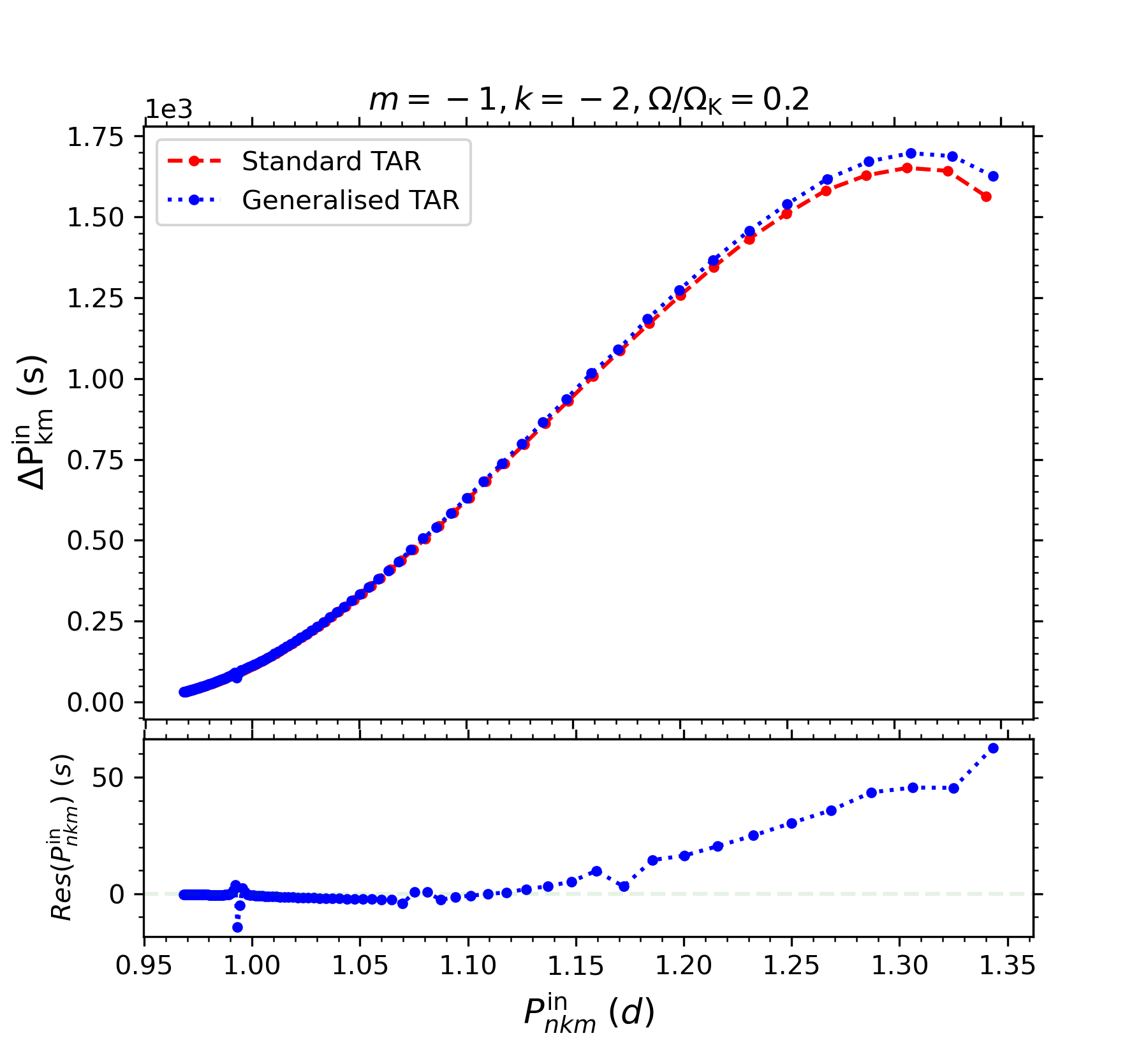}}
    \caption{Period spacing pattern in the inertial frame  for $\{k=0,m=1\}$ (above) and $\{k=-2,m=-1\}$ (below) modes at $\Omega/\Omega_{\rm K}=20\%$ within the standard (red line) and the generalised (blue line) TAR frameworks. The bottom panel for each mode shows the differences with respect to the standard TAR period spacing pattern (the sudden fluctuations in the period spacing pattern are caused by the numerical noise which is introduced by the numerical derivatives of the mapping with respect to $\zeta$ and $\theta$ used in the resolution of the GLTE).}
    \label{fig:periodspacing}
\end{figure}

To compute the period spacing patterns, we first calculated  the asymptotic frequencies by following the method developed by \cite{Henneco2021} (the detailed steps of this computation can be found in their Appendix\;B). First, we compute for each radial order $n$ of a given mode $(k,m)$, the corresponding $\nu_{n k m}=2\Omega/\omega_{nkm}$ using the implicit equation of the asymptotic frequencies of low-frequency GIWs (Eq.\;\ref{eq:frequencies}). Then, by taking their inverse and multiplying them with $2\Omega$, we retrieve the asymptotic frequencies $\omega_{nkm}$. The corresponding asymptotic frequencies in the inertial (observer's) frame $\omega_{nkm}^{\mathrm{in}}$ are then found using the angular Doppler shift (Eq.\;\ref{eq:doppler_shift})
\begin{equation}
    \omega_{nkm}^{\mathrm{in}} = m \Omega + \underbrace{ \omega_{nkm}}_{=2\Omega / \nu_{n k m} }.
\end{equation}
Now, we can easily calculate the asymptotic periods  in the inertial frame ($P_{nkm}^{\mathrm{in}}=2\pi/ \omega_{nkm}^{\mathrm{in}}$) and the corresponding period spacing which is shown in Fig.\;\ref{fig:periodspacing} for the prograde dipole sectoral $\{k=0,m=1\}$ and the retrograde Rossby $\{k=-2,m=-1\}$ modes using the standard TAR (in spherical geometry) and the generalised TAR. The asymptotic periods within the standard TAR are computed using the following asymptotic frequencies deduced from the SLTE (Eq.\;\ref{eq:slte}) 
\begin{equation}
\omega_{n k m}^{\rm stand}=\frac{\Lambda_{\nu k m}^{1 / 2}}{(n+1 / 2) \pi}\int_{\zeta_{1}}^{\zeta_{2}} \frac{ N(\zeta)}{\zeta} \mathrm{d} \zeta.
\end{equation}

By  comparing  the  period spacing pattern computed using the standard TAR and the generalised TAR in each of bottom panels of Fig.\;\ref{fig:periodspacing}, we find that the spacing values of the $\{k=0,m=1\}$ and $\{k=-2,m=-1\}$ modes increase with the decrease in radial orders (low (high) radial orders equivalent to short (long) pulsation periods for $\rm g$ modes and long (short) pulsation periods for $\rm r$ modes) under  the influence of the centrifugal deformation. The effect of the centrifugal acceleration on the period spacing that we found here is different from the one discussed in \cite{Henneco2021} who found an increase in the period spacing of the two modes.
Another notable difference between the generalised TAR and the standard \citep{lee+saio1997} and the perturbative \citep{Henneco2021} TAR is in the domain of validity and applicability, where the general treatment adopted in this work should be more reliable than the standard and the perturbative ones.  Fig.\;\ref{fig:periodspacing} reveals the difference in the period spacing values computed with the standard and the generalised formalisms and also offers a comparison of this difference with typical observational uncertainties of measured period spacing values from modern space photometric observations. In fact, \citet{vanreeth2015} shows that for a sample of 40 $\gamma\,$Dor stars the error margins of the period spacings have values up to 25 seconds for the prograde dipole modes (larger than our period spacing differences) and between 20 and 50 seconds for the retrograde Rossby modes (smaller than our  period spacing differences).

\subsection{Detectability of the centrifugal deformation effect}
\begin{figure}
    \centering
     \resizebox{\hsize}{!}{\includegraphics{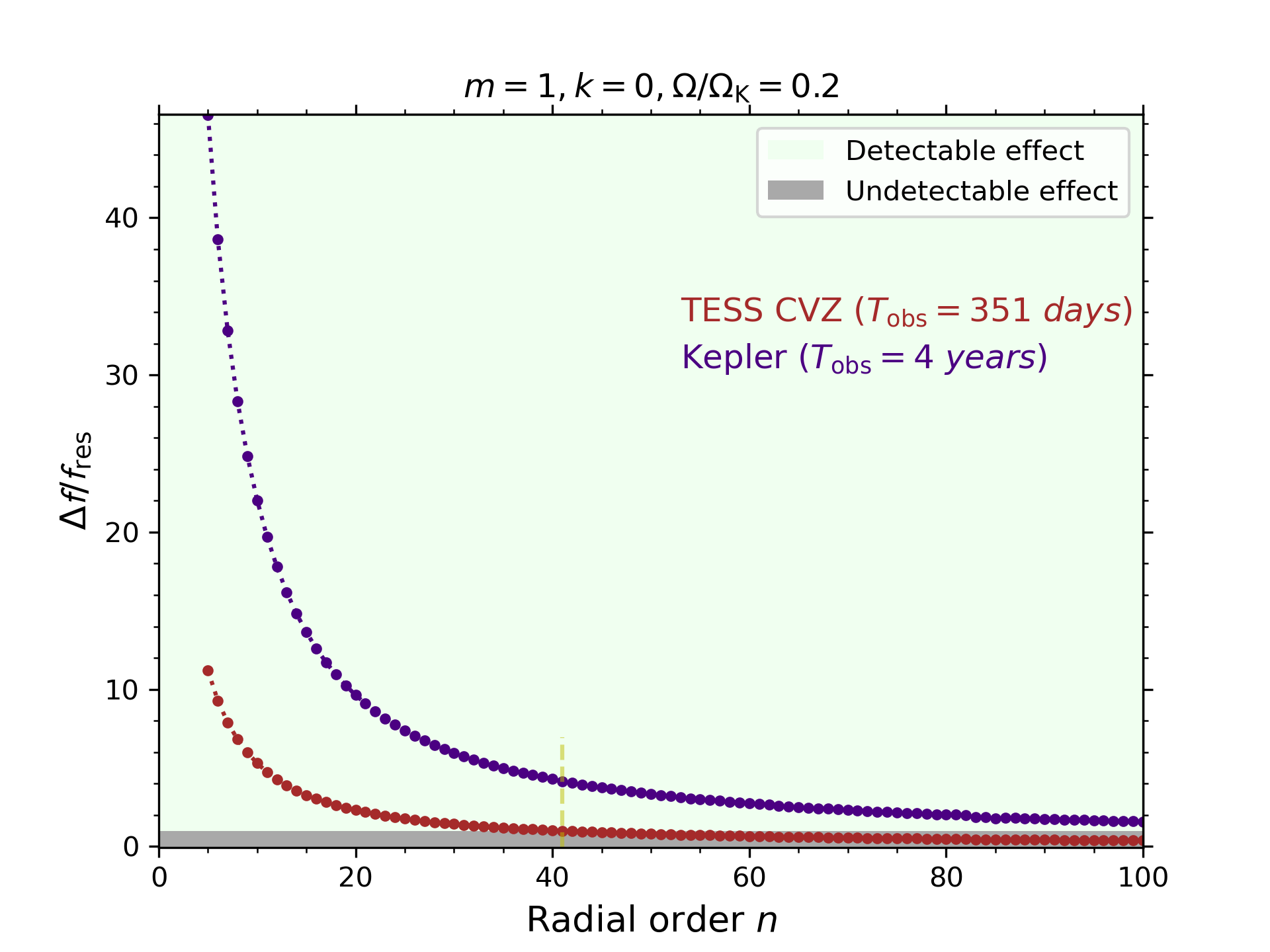}}
    \resizebox{\hsize}{!}{\includegraphics{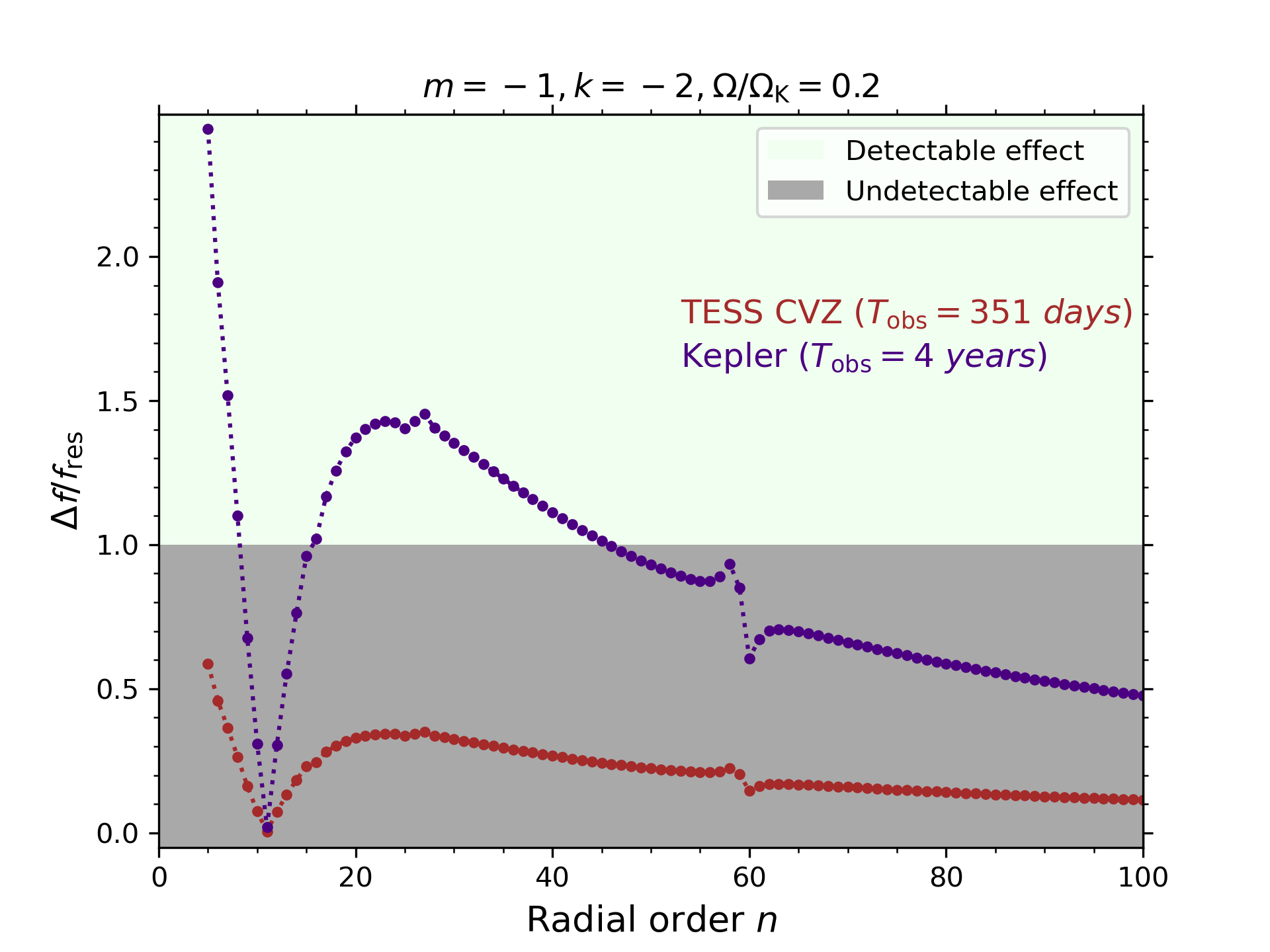}}
    \caption{Detectability of $\{k=0,m=1\}$ (above) and $\{k=-2,m=-1\}$ (below) modes as a function of the radial order $n$ at a rotation rate $\Omega/\Omega_{\mathrm{\rm K}}=20\%$ based on the  frequency resolution of TESS (brown line) and \textit{Kepler} (indigo line).}
    \label{fig:detectability}
\end{figure}

To evaluate the detectability of the centrifugal effect with space-based photometric observations, we first compute  the frequency differences $\Delta f_{\rm centrifugal}$ between asymptotic frequencies calculated in the standard TAR ($\rm TAR_{\rm s}$) and those computed in the generalised TAR ($\rm TAR_{\rm g}$) through Eq.\,(\ref{eq:frequencies}):
\begin{equation}
    \Delta f_{\rm centrifugal}(n) = |f_{\rm TAR_{\rm g}}(n) - f_{\rm TAR_{\rm s}}(n) |.
\end{equation}
Then, by comparing the obtained frequency differences with the frequency resolutions ($f_{\mathrm{res}}=1/T_{\mathrm{obs}}$) of nominal 4-year \textit{Kepler} \citep{borucki2009} and TESS CVZ \citep[Transiting Exoplanet Survey Satellite Continuous Viewing Zone;][]{ricker2014} light curves covering  quasi-continuously observation times of $T_{\mathrm{obs}} = 4\,$years and $T_{\mathrm{obs}} = 351\,$days, respectively, we can deduce the radial orders $n$ for which the frequency differences are, at least in principle, expected to be detectable. In Fig.\,\ref{fig:detectability}, we display these results for the $\{k=0,m=1\}$ and $\{k=-2,m=-1\}$ modes for a stellar model rotating at $0.2\,\Omega_{\mathrm{\rm K}}$. We can see that the centrifugal effect for the $\{k=0,m=1\}$ mode is detectable for all radial orders using \textit{Kepler} observations, but detectable only for $n< 41$ using TESS observations (yellow dashed line in Fig.\,\ref{fig:detectability}). Indeed even if for $n\in[20,40]$ using nominal TESS CVZ light curves and  $n\in[75,100]$ using nominal \textit{Kepler} light curves  we have $\Delta f_{\rm centrifugal}/f_{\rm res}\gtrsim 1$, so we would expect that the detectability of the centrifugal effect for these radial orders is in principle guaranteed. However, since the frequency differences is nearly equal to the resolution one ($\Delta f_{\rm centrifugal}\approx f_{\rm res}$), this effect  become difficult to detect in this case.  Regarding the $\{k=-2,m=-1\}$ mode, we can see that the centrifugal effect is less detectable in comparison with the $\{k=0,m=1\}$ mode but remains in principle observable (yet difficult to observe because $\Delta f_{\rm centrifugal}\approx f_{\rm res}$) using \textit{Kepler} observations for small radial orders ($n\in[5,8]$) and for $n\in[16,45]$.

Our results in Figs.\,\ref{fig:periodspacing}\,\&\,\ref{fig:detectability}  are formal comparisons in the sense that uncalibrated astrophysical effects occurring in stellar models (e.g. the presence of rotational mixing and atomic diffusion effects) usually imply larger uncertainties in the prediction of the period spacing values than the differences shown in these figures, as outlined in \cite{aerts2018}. For this reason, our theoretical findings in the current work offer a key result in terms of reliability of future asteroseismic analyses of observed pulsators. The centrifugal effect is not significant when assessed within the validity domain of the generalised TAR which is consistent with the results of \cite{ouazzani2017} and \cite{Henneco2021}.

\section{Evaluation of the terms hierarchy imposed by the TAR within uniformly rotating deformed stars}\label{sect:hierarchy_validation}
The hierarchy of terms imposed by the TAR can be summarised by the following frequency hierarchy:
\begin{gather}
    2\Omega\ll N,\label{eq:hier1}\\
    \omega\ll N,\label{eq:hier2}
\end{gather}
which ensures the other hierarchies. In fact, the vertical wave vector is larger than the horizontal one ($|k_H|\ll |k_V|$) which comes from the dispersion relation of GIWs \citep{lee+saio1997}
    \begin{equation}\label{des}
    \omega^{2} \approx \frac{N^{2} k_{H}^{2}+(2 \vec{\Omega} \cdot \vec{k})^{2}}{k^{2}}.
    \end{equation}
    In addition, the latitudinal component of the rotation vector in the momentum equation $\Omega_{\mathrm{H}}=\Omega \sin \theta$ is neglected for all colatitudes because $(2\vec{\Omega} \cdot \vec{k}) \approx 2\Omega_V k_V$. And finally, the wave velocity is almost horizontal ($|v^\zeta|\ll\{|v^\theta|,|v^\varphi|\}$). Indeed, by adopting the anelastic approximation we obtain
    \begin{align}
    \vec{\nabla} \cdot(\rho \vec{v})\approx0 &\Rightarrow \vec{k} \cdot \vec{v}= k_{V} v_{V}+\vec{k}_{H} \cdot \vec{v}_H\approx0 \nonumber\\ &\Rightarrow \frac{v_V}{v_H}\approx-\frac{k_H}{k_V}\ll1.
\end{align}
The anelastic approximation filters out acoustic waves which have higher frequencies, and is justified by the fact that we study only low frequency waves (Eq.\;\ref{eq:hier2}) within the framework of the TAR treatment.

So in order to verify whether the TAR is still valid in uniformly rotating deformed stars, we  discuss the frequency hierarchy (Eqs.\;\ref{eq:hier1}\;\&\;\ref{eq:hier2}) using the Brunt–Väisälä frequency and rotation profiles from ESTER models and using the asymptotic frequencies calculated in Sect.\;\ref{sect:seismic_diagnosis}.

\subsection{The strong stratification assumption: $2\Omega\ll N$}
\begin{figure}
    \centering
    \resizebox{\hsize}{!}{\includegraphics{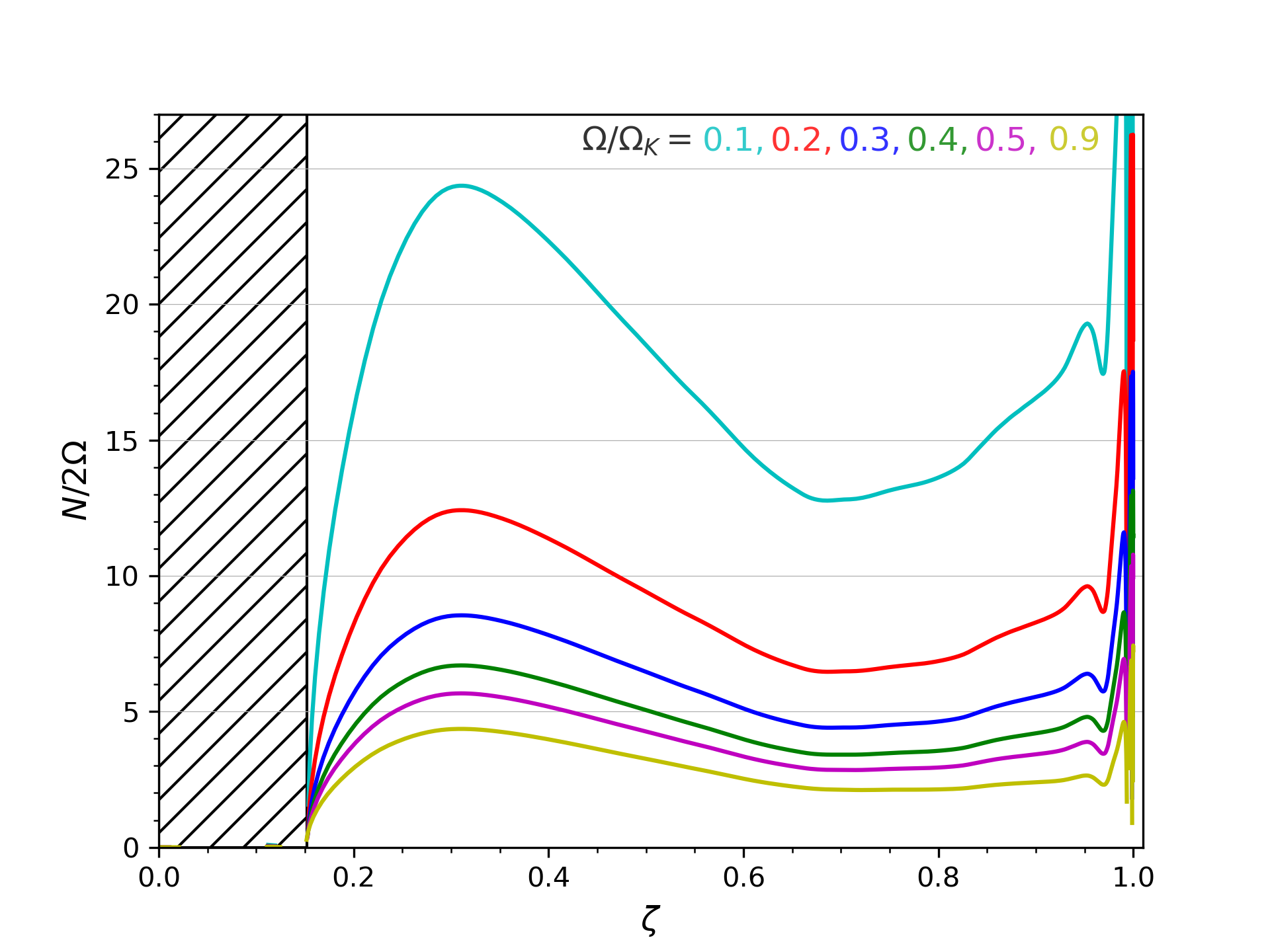}}
    \caption{$N/2\Omega$ term as a function of the pseudo-radius $\zeta$ at different rotation rates $\Omega/\Omega_{\rm K}$ (the hatched area represents the convective region of the star).}
    \label{fig:tarverifrot}
\end{figure}

We evaluate the term $N/2\Omega$ using the Brunt–Väisälä frequency and the angular velocity computed with ESTER models  ($3\,\mathrm{M}_{\odot}$, $X_{\mathrm{c}}=0.7$) for rotation rates $\Omega/\Omega_{\mathrm{\rm K}} \in \left[0.1, 0.9\right]$. As shown in Fig.\,\ref{fig:tarverifrot}, the strong stratification assumption is valid only in the radiative zone far from the border between convection and radiation ($\zeta \ge 0.2 $ at $\Omega/\Omega_{\mathrm{\rm K}}\le 0.2$). Beyond this critical value, the 
Brunt–Väisälä frequency and the frequency of rotation have a very close order of magnitude so we can no longer adopt the strong stratification approximation for stars close to the zero-age main-sequence. As stars evolve, however, they build up a strong gradient of molecular weight, increasing the value of $N(\zeta)$ in the region near their convective core. Therefore, the strong stratification assumption becomes better fulfilled as stars evolve along the main sequence because the spin-up of their core remains modest \citep{Li2020,aerts2021}, while $N(\zeta)$ increases.

\subsection{The low frequency assumption: $\omega\ll N$}
\begin{figure}
    \centering
    \resizebox{\hsize}{!}{\includegraphics{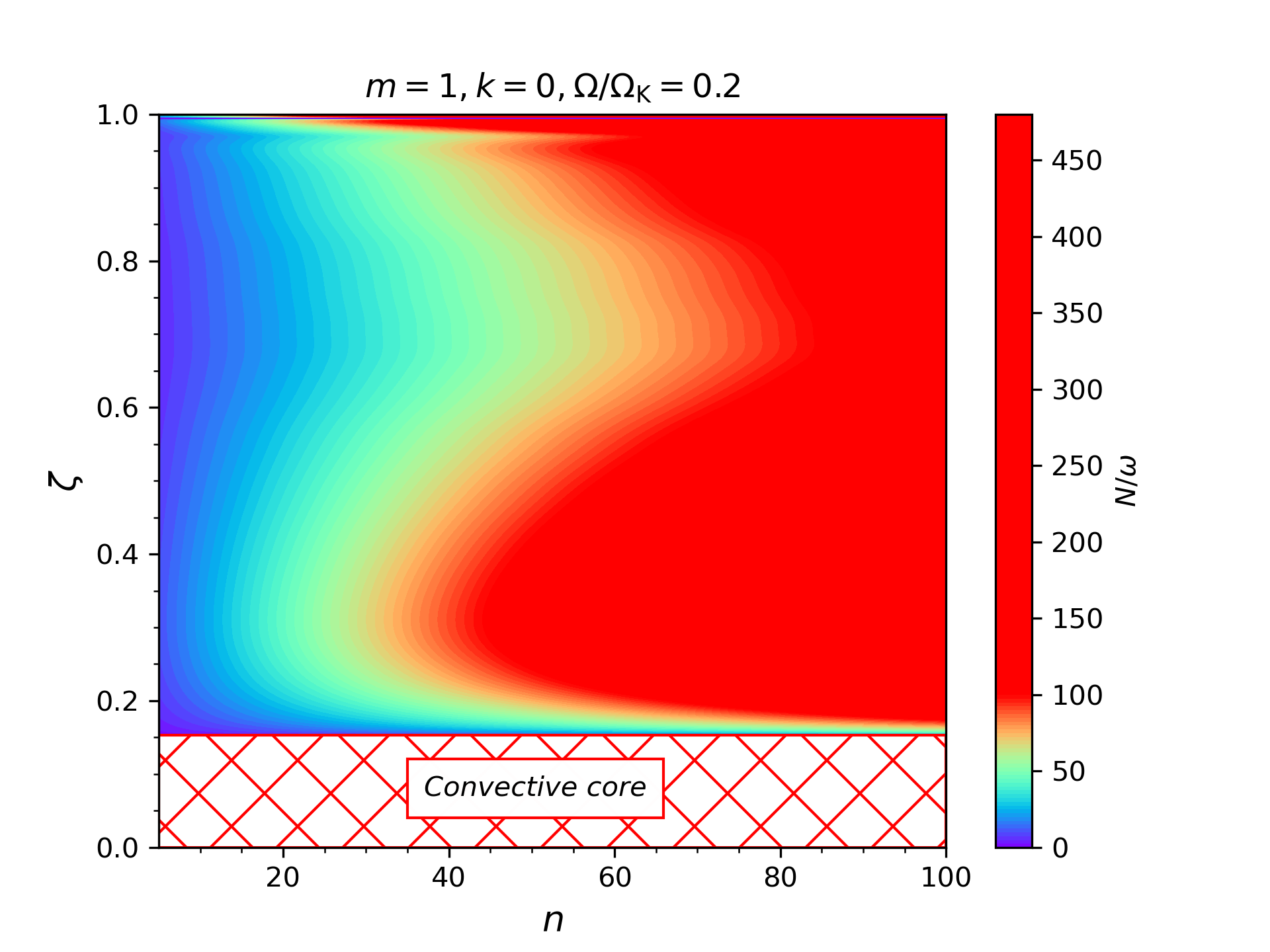}}
    \resizebox{\hsize}{!}{\includegraphics{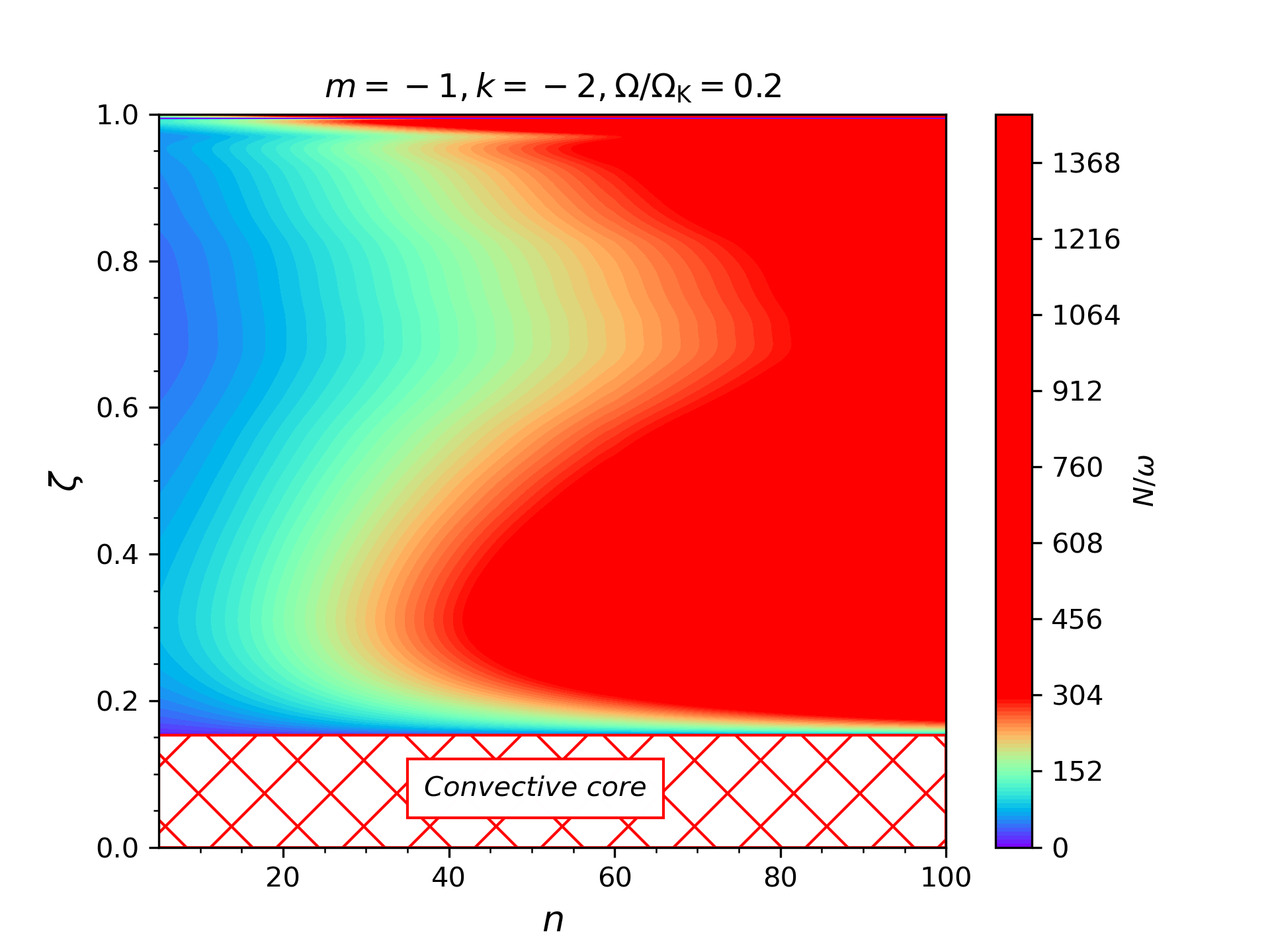}}
    \caption{$N/\omega$ term as a function of the pseudo-radius $\zeta$ and the radial order $n$ for $\{k=0,m=1\}$ (top panel) and $\{k=-2,m=-1\}$ (bottom panel) modes at $\Omega/\Omega_{\rm K}=20\%$.}
    \label{fig:tarveriffreq}
\end{figure}

We evaluate the term $N/\omega$  using the Brunt–Väisälä frequency computed by ESTER models and the asymptotic frequencies for $\{k=0,m=1\}$ and $\{k=-2,m=-1\}$ modes calculated in Sect.\;\ref{sect:seismic_diagnosis}. As shown in Fig.\,\ref{fig:tarveriffreq}, the low frequency assumption  is valid for $\{k=-2,m=-1\}$ mode for all radial orders and pseudo-radii while for  $\{k=0,m=1\}$ mode this assumption  is valid for all radial orders only if we are far from the transition layer between the convective core and the radiative envelope. If this is not the case, we have to work with high radial orders ($n>20$). Below this critical value, the Brunt–Väisälä frequency and the wave frequency have a very close order of magnitude so we can no longer adopt the low frequency approximation.

\section{Discussion and conclusions}\label{sect:conclusion}
The aim of this work was to study the possibility of carrying out a new generalisation of the TAR that takes into account the centrifugal acceleration in the case of strongly deformed, rapidly rotating stars and planets. We approached this issue by deriving the generalised Laplace Tidal equation in a general spheroidal coordinate system. We relied on two necessary assumptions (described by Eqs.\;\ref{eq:approx1} and \ref{eq:approx2}) that are applicable in a specific domain, which will then define the validity domain of the generalised TAR. The equation that we derive has a similar  form to the one obtained when the TAR is applied to weakly rotating spherical stars \citep{lee+saio1997}, differentially rotating spherical stars \citep{mathis2009,vanreeth2018} and moderately rotating, weakly deformed stars \citep{mathis+prat2019}. This new formalism can be applied in the radiative region of all types of stars and planets.

Using this theoretical framework with 2D ESTER models we can define the validity domain of the TAR with a chosen confidence level as a function of the rotation rate and the pseudo-radius. We apply this general formalism to rapidly rotating early-type stars $(\mathrm{M}=3 \mathrm{M}_{\odot},~X_{\rm c} = 0.7)$ , where we found that the generalised TAR is only applicable for all rotation rates for values of the pseudo-radius smaller than $0.57$ ($\zeta_{\rm limit}<0.57$). To calculate the asymptotic pulsation frequencies we then have to integrate the dispersion relation on the pulsation mode cavity (the radiative zone) (Eq.\;\ref{eq:quantisation}) where $\zeta \in [\zeta_{1},\zeta_{2}]$ (in our considered case $\zeta_1=0.153$ and $\zeta_2=1$). This condition will limit the rotation rates where the TAR is valid. We can therefore deduce a rotation rate limit ($\Omega_{\rm limit}=0.2\Omega_{\rm K}$ in our case) where the TAR ceases to be applicable with a fixed level of confidence ($90\%$ in our study).

We found a difference between the limit of the TAR applicability defined in \citet{mathis+prat2019} which was $\Omega_{\rm limit}=0.4\Omega_{\rm K}$ and the one defined here. The current new limit should be more realistic and accurate since it is first derived from a 2D model which takes the impact of the centrifugal acceleration into account in a non-perturbative way. Next, the validity domain of the adopted approximations is studied carefully in more details and with a level of confidence of $90\%$.

The centrifugal acceleration affects two aspects within our new generalised theoretical formalism compared to the standard application of the TAR \citep{lee+saio1997}. On the one hand, the stellar structure as a whole becomes deformed which causes the usual radial coordinate $r$, the hydrostatic background and thus the Brunt-Vaïsälä frequency to have a dependency not only on the pseudo-radius $\zeta$ but also on the colatitude $\theta$. This introduces a new challenge in the decoupling of the vertical and the horizontal dynamics and the applicability of the quantisation relation unless we adopt adequate approximations which will  restrict the validity domain of our formalism. On the other hand, the dynamics of the GIWs represented by the GLTE, is altered and get a radial dependence compared to the one in the spherical case represented by the SLTE. Then, we apply our new formalism to a rapidly rotating early-type star $(\mathrm{M}=3 \mathrm{M}_{\odot},~X_{\rm c} = 0.7,~\Omega/\Omega_{\rm K}=20\%)$. In this case, we observe a non-monotonous geometrical behaviour in the Rossby-like solutions and a monotonous one (only an inward migration) in the gravity-like solutions. In fact, near the equator the Rossby-like solutions migrate onwards, causing a broadening of its shape. But far from the equator this behaviour becomes less and less visible until it changes to the opposite one where these solutions migrate inwards, causing a narrowing of its shape. We found also that the centrifugal acceleration increases the  period spacing values for low radial orders (equivalent to short pulsation periods for $\rm g$ modes and long pulsation periods for $\rm r$ modes) and decrease slightly for high radial orders (equivalent to long pulsation periods for $\rm g$ modes and short pulsation periods for $\rm r$ modes). But yet the effects of the centrifugal acceleration (within the validity domain of the generalised TAR) are of the order of or smaller than the effects of other missing stellar physical processes in our models (e.g. accurate rotational mixing, atomic diffusion, etc.). This can mask the effect of the centrifugal acceleration in forward asteroseismic modelling analyses, even when it is theoretically detectable.

This new generalisation of the TAR can also be used to study the tidal dissipation induced by low-frequency GIWs in rapidly rotating deformed stars and planets \citep{braviner2014} and the angular momentum transport with a formalism directly implantable in ESTER models. In addition, we can apply our formalism to different 2D ESTER model grids and then compare the obtained results with direct computations using 2D oscillation codes TOP \citep{Reese2021} and ACOR \citep{ouazzani2017}.
\;\newline \;

The next steps will be first to take into account the general differential rotation in the deformed case. So far, \citet{ogilvie+lin2004} and \citet{mathis2009} included the effects of general differential rotation on low-frequency GIWs within the framework of the TAR for spherical stars. Then, their formalism has been successfully applied in \citet{vanreeth2018} to derive the variation of the asymptotic period spacing in the case of a weak radial differential rotation and evaluate the sensitivity of GIWs to the effect of differential rotation in spherical stars. Next, we will take into account the magnetic field in a non-perturbative way \citep{Mathis+deBrye2011,Mathis+deBrye2012}. So far, \cite{prat2019,prat2020} and \cite{VanBeeck2020} have focused on the case where magnetic fields are weak enough to be treated within a perturbative treatment to study the effects of a magnetic field on the seismic parameters of $\rm g$ modes which become magneto-gravito inertial modes.

So in the near future, magneto-gravito inertial waves have to be studied in the general case of rapidly and differentially rotating stars and planets. As a first step, in a forthcoming paper (Paper II, submitted) we will examine if the TAR can be generalised to the case of strongly deformed differentially rotating stars and planets.

\begin{acknowledgements}
We thank the referee, Pr.~H.Shibahashi, for his constructive report that allowed us to improve the quality of our work and of the article. We thank the ESTER code developers for their efforts and for making their codes publicly available. We are also grateful to Pr.~C.Aerts who kindly commented on the manuscript and suggested improvements. H.D., V.P. and S.M. acknowledge support from the European Research Council through ERC grant SPIRE 647383.
 H.D. and S.M. acknowledge support from the CNES PLATO grant at CEA/DAp. TVR gratefully acknowledges support from the Research Foundation Flanders (FWO) under grant agreement No. 12ZB620N.
\end{acknowledgements}

\bibliographystyle{aa}
\bibliography{bibliography}

\begin{thebibliography}{67}
\expandafter\ifx\csname natexlab\endcsname\relax\def\natexlab#1{#1}\fi

\bibitem[{Aerts(2021)}]{aerts2021}
Aerts, C. 2021, Rev. Mod. Phys., 93, 015001

\bibitem[{{Aerts} {et~al.}(2010){Aerts}, {Christensen-Dalsgaard}, \&
  {Kurtz}}]{aerts2010}
{Aerts}, C., {Christensen-Dalsgaard}, J., \& {Kurtz}, D.~W. 2010,
  {Asteroseismology} (Springer)

\bibitem[{{Aerts} {et~al.}(2019){Aerts}, {Mathis}, \& {Rogers}}]{aerts2019}
{Aerts}, C., {Mathis}, S., \& {Rogers}, T.~M. 2019, \araa, 57, 35

\bibitem[{{Aerts} {et~al.}(2018){Aerts}, {Molenberghs}, {Michielsen},
  {Pedersen}, {Bj{\"o}rklund}, {Johnston}, {Mombarg}, {Bowman}, {Buysschaert},
  {P{\'a}pics}, {Sekaran}, {Sundqvist}, {Tkachenko}, {Truyaert}, {Van Reeth},
  \& {Vermeyen}}]{aerts2018}
{Aerts}, C., {Molenberghs}, G., {Michielsen}, M., {et~al.} 2018, \apjs, 237, 15

\bibitem[{{Berthomieu} {et~al.}(1978){Berthomieu}, {Gonczi}, {Graff},
  {Provost}, \& {Rocca}}]{Berthomieu1978}
{Berthomieu}, G., {Gonczi}, G., {Graff}, P., {Provost}, J., \& {Rocca}, A.
  1978, \aap, 70, 597

\bibitem[{{Bildsten} {et~al.}(1996){Bildsten}, {Ushomirsky}, \&
  {Cutler}}]{Bildsten1996}
{Bildsten}, L., {Ushomirsky}, G., \& {Cutler}, C. 1996, \apj, 460, 827

\bibitem[{Bonazzola {et~al.}(1998)Bonazzola, Gourgoulhon, \&
  Marck}]{bonazzola1998}
Bonazzola, S., Gourgoulhon, E., \& Marck, J.-A. 1998, Phys. Rev. D, 58, 104020

\bibitem[{{Borucki} {et~al.}(2009){Borucki}, {Koch}, {Batalha}, {Caldwell},
  {Christensen-Dalsgaard}, {Cochran}, {Dunham}, {Gautier}, {Geary},
  {Gilliland}, {Jenkins}, {Kjeldsen}, {Lissauer}, \& {Rowe}}]{borucki2009}
{Borucki}, W., {Koch}, D., {Batalha}, N., {et~al.} 2009, in IAU Symposium, Vol.
  253, Transiting Planets, ed. F.~{Pont}, D.~{Sasselov}, \& M.~J. {Holman},
  289--299

\bibitem[{{Bouabid} {et~al.}(2013){Bouabid}, {Dupret}, {Salmon},
  {Montalb{\'a}n}, {Miglio}, \& {Noels}}]{bouabid2013}
{Bouabid}, M.~P., {Dupret}, M.~A., {Salmon}, S., {et~al.} 2013, \mnras, 429,
  2500

\bibitem[{{Braviner} \& {Ogilvie}(2014)}]{braviner2014}
{Braviner}, H.~J. \& {Ogilvie}, G.~I. 2014, \mnras, 441, 2321

\bibitem[{Christensen-Dalsgaard(1997)}]{Christensen1997}
Christensen-Dalsgaard, J. 1997, Stellar Oscillations

\bibitem[{{Cowling}(1941)}]{cowling1941}
{Cowling}, T.~G. 1941, \mnras, 101, 367

\bibitem[{{Dintrans} \& {Rieutord}(2000)}]{Dintrans+Rieutord2000}
{Dintrans}, B. \& {Rieutord}, M. 2000, \aap, 354, 86

\bibitem[{{Dintrans} {et~al.}(1999){Dintrans}, {Rieutord}, \&
  {Valdettaro}}]{Dintrans1999}
{Dintrans}, B., {Rieutord}, M., \& {Valdettaro}, L. 1999, Journal of Fluid
  Mechanics, 398, 271

\bibitem[{{Eckart}(1960)}]{eckart1960}
{Eckart}, C. 1960, {Hydrodynamics of Oceans and Atmospheres} (Pergamon Press
  (Oxford))

\bibitem[{{Espinosa Lara} \& {Rieutord}(2013)}]{espinosa+rieutord2013}
{Espinosa Lara}, F. \& {Rieutord}, M. 2013, A\&A, 552, A35

\bibitem[{Fr{\"o}man \& Fr{\"o}man(1965)}]{froman1965}
Fr{\"o}man, N. \& Fr{\"o}man, P.~O. 1965, JWKB approximation : contributions to
  the theory (Amsterdam: North-Holland Publishing Company)

\bibitem[{Gough(1993)}]{gough1993}
Gough, D. 1993, Les Houches Session XLVII, ed. J.-P. Zahn, \& J. Zinn-Justin
  (Amsterdam: Elsevier), 399

\bibitem[{{Henneco} {et~al.}(2021){Henneco}, {Van Reeth}, {Prat}, {Mathis},
  {Mombarg}, \& {Aerts}}]{Henneco2021}
{Henneco}, J., {Van Reeth}, T., {Prat}, V., {et~al.} 2021, arXiv e-prints,
  arXiv:2101.04116

\bibitem[{{Hough}(1898)}]{hough1898}
{Hough}, S.~S. 1898, Philosophical Transactions of the Royal Society of London
  Series A, 191, 139

\bibitem[{{Laplace}(1799)}]{laplace1799}
{Laplace}, P.~S. 1799, {Trait\'e de M\'ecanique C\"eleste} (Imprimerie de
  Crapelet (Paris))

\bibitem[{{Lee}(2019)}]{lee2019}
{Lee}, U. 2019, \mnras, 484, 5845

\bibitem[{{Lee} \& {Baraffe}(1995)}]{lee+baraffe1995}
{Lee}, U. \& {Baraffe}, I. 1995, \aap, 301, 419

\bibitem[{{Lee} {et~al.}(2014){Lee}, {Neiner}, \& {Mathis}}]{lee2014}
{Lee}, U., {Neiner}, C., \& {Mathis}, S. 2014, \mnras, 443, 1515

\bibitem[{{Lee} \& {Saio}(1993)}]{lee+saio1993}
{Lee}, U. \& {Saio}, H. 1993, \mnras, 261, 415

\bibitem[{{Lee} \& {Saio}(1997)}]{lee+saio1997}
{Lee}, U. \& {Saio}, H. 1997, \apj, 491, 839

\bibitem[{{Lee} \& {Saio}(2020)}]{lee+saio2020}
{Lee}, U. \& {Saio}, H. 2020, \mnras, 497, 4117

\bibitem[{{Li} {et~al.}(2019{\natexlab{a}}){Li}, {Bedding}, {Murphy}, {Van
  Reeth}, {Antoci}, \& {Ouazzani}}]{li2019a}
{Li}, G., {Bedding}, T.~R., {Murphy}, S.~J., {et~al.} 2019{\natexlab{a}},
  \mnras, 482, 1757

\bibitem[{{Li} {et~al.}(2019{\natexlab{b}}){Li}, {Van Reeth}, {Bedding},
  {Murphy}, \& {Antoci}}]{li2019b}
{Li}, G., {Van Reeth}, T., {Bedding}, T.~R., {Murphy}, S.~J., \& {Antoci}, V.
  2019{\natexlab{b}}, \mnras, 487, 782

\bibitem[{{Li} {et~al.}(2020){Li}, {Van Reeth}, {Bedding}, {Murphy}, {Antoci},
  {Ouazzani}, \& {Barbara}}]{Li2020}
{Li}, G., {Van Reeth}, T., {Bedding}, T.~R., {et~al.} 2020, \mnras, 491, 3586

\bibitem[{{Ligni{\`e}res} {et~al.}(2006){Ligni{\`e}res}, {Rieutord}, \&
  {Reese}}]{lignieres2006}
{Ligni{\`e}res}, F., {Rieutord}, M., \& {Reese}, D. 2006, \aap, 455, 607

\bibitem[{{Mathis}(2009)}]{mathis2009}
{Mathis}, S. 2009, A\&A, 506, 811

\bibitem[{{Mathis} \& {de Brye}(2011)}]{Mathis+deBrye2011}
{Mathis}, S. \& {de Brye}, N. 2011, \aap, 526, A65

\bibitem[{{Mathis} \& {de Brye}(2012)}]{Mathis+deBrye2012}
{Mathis}, S. \& {de Brye}, N. 2012, \aap, 540, A37

\bibitem[{{Mathis} \& {Prat}(2019)}]{mathis+prat2019}
{Mathis}, S. \& {Prat}, V. 2019, A\&A, 631, A26

\bibitem[{{Mathis} {et~al.}(2008){Mathis}, {Talon}, {Pantillon}, \&
  {Zahn}}]{mathis2008}
{Mathis}, S., {Talon}, S., {Pantillon}, F.~P., \& {Zahn}, J.~P. 2008, \solphys,
  251, 101

\bibitem[{{Neiner} {et~al.}(2012){Neiner}, {Floquet}, {Samadi}, {Espinosa
  Lara}, {Fr{\'e}mat}, {Mathis}, {Leroy}, {de Batz}, {Rainer}, {Poretti},
  {Mathias}, {Guarro Fl{\'o}}, {Buil}, {Ribeiro}, {Alecian}, {Andrade},
  {Briquet}, {Diago}, {Emilio}, {Fabregat}, {Guti{\'e}rrez-Soto}, {Hubert},
  {Janot-Pacheco}, {Martayan}, {Semaan}, {Suso}, \& {Zorec}}]{neiner2012}
{Neiner}, C., {Floquet}, M., {Samadi}, R., {et~al.} 2012, \aap, 546, A47

\bibitem[{{Neiner} {et~al.}(2020){Neiner}, {Lee}, {Mathis}, {Saio}, {Lovekin},
  \& {Augustson}}]{Neiner2020}
{Neiner}, C., {Lee}, U., {Mathis}, S., {et~al.} 2020, \aap, 644, A9

\bibitem[{{Ogilvie} \& {Lin}(2004)}]{ogilvie+lin2004}
{Ogilvie}, G.~I. \& {Lin}, D.~N.~C. 2004, \apj, 610, 477

\bibitem[{{Ogilvie} \& {Lin}(2007)}]{ogilvie+lin2007}
{Ogilvie}, G.~I. \& {Lin}, D.~N.~C. 2007, \apj, 661, 1180

\bibitem[{{Ouazzani} {et~al.}(2012){Ouazzani}, {Dupret}, \&
  {Reese}}]{Ouazzani2012}
{Ouazzani}, R.~M., {Dupret}, M.~A., \& {Reese}, D.~R. 2012, \aap, 547, A75

\bibitem[{{Ouazzani} {et~al.}(2019){Ouazzani}, {Marques}, {Goupil},
  {Christophe}, {Antoci}, {Salmon}, \& {Ballot}}]{ouazzani2019}
{Ouazzani}, R.~M., {Marques}, J.~P., {Goupil}, M.~J., {et~al.} 2019, \aap, 626,
  A121

\bibitem[{{Ouazzani} {et~al.}(2017){Ouazzani}, {Salmon}, {Antoci}, {Bedding},
  {Murphy}, \& {Roxburgh}}]{ouazzani2017}
{Ouazzani}, R.-M., {Salmon}, S.~J.~A.~J., {Antoci}, V., {et~al.} 2017, \mnras,
  465, 2294

\bibitem[{{Pantillon} {et~al.}(2007){Pantillon}, {Talon}, \&
  {Charbonnel}}]{pantillon2007}
{Pantillon}, F.~P., {Talon}, S., \& {Charbonnel}, C. 2007, \aap, 474, 155

\bibitem[{{P{\'a}pics} {et~al.}(2015){P{\'a}pics}, {Tkachenko}, {Aerts}, {Van
  Reeth}, {De Smedt}, {Hillen}, {{\O}stensen}, \& {Moravveji}}]{Papics2015}
{P{\'a}pics}, P.~I., {Tkachenko}, A., {Aerts}, C., {et~al.} 2015, \apjl, 803,
  L25

\bibitem[{{P{\'a}pics} {et~al.}(2017){P{\'a}pics}, {Tkachenko}, {Van Reeth},
  {Aerts}, {Moravveji}, {Van de Sande}, {De Smedt}, {Bloemen}, {Southworth},
  {Debosscher}, {Niemczura}, \& {Gameiro}}]{Papics2017}
{P{\'a}pics}, P.~I., {Tkachenko}, A., {Van Reeth}, T., {et~al.} 2017, \aap,
  598, A74

\bibitem[{{Pedersen} {et~al.}(2018){Pedersen}, {Aerts}, {P{\'a}pics}, \&
  {Rogers}}]{pedersen2018}
{Pedersen}, M.~G., {Aerts}, C., {P{\'a}pics}, P.~I., \& {Rogers}, T.~M. 2018,
  \aap, 614, A128

\bibitem[{{Prat} {et~al.}(2019){Prat}, {Mathis}, {Buysschaert}, {Van Beeck},
  {Bowman}, {Aerts}, \& {Neiner}}]{prat2019}
{Prat}, V., {Mathis}, S., {Buysschaert}, B., {et~al.} 2019, \aap, 627, A64

\bibitem[{{Prat} {et~al.}(2020){Prat}, {Mathis}, {Neiner}, {Van Beeck},
  {Bowman}, \& {Aerts}}]{prat2020}
{Prat}, V., {Mathis}, S., {Neiner}, C., {et~al.} 2020, \aap, 636, A100

\bibitem[{{Reese} {et~al.}(2006){Reese}, {Ligni{\`e}res}, \&
  {Rieutord}}]{reese2006}
{Reese}, D., {Ligni{\`e}res}, F., \& {Rieutord}, M. 2006, \aap, 455, 621

\bibitem[{{Reese} {et~al.}(2009){Reese}, {MacGregor}, {Jackson}, {Skumanich},
  \& {Metcalfe}}]{reese2009}
{Reese}, D.~R., {MacGregor}, K.~B., {Jackson}, S., {Skumanich}, A., \&
  {Metcalfe}, T.~S. 2009, A\&A, 506, 189

\bibitem[{{Reese} {et~al.}(2021){Reese}, {Mirouh}, {Espinosa Lara}, {Rieutord},
  \& {Putigny}}]{Reese2021}
{Reese}, D.~R., {Mirouh}, G.~M., {Espinosa Lara}, F., {Rieutord}, M., \&
  {Putigny}, B. 2021, \aap, 645, A46

\bibitem[{{Ricker} {et~al.}(2014){Ricker}, {Winn}, {Vanderspek}, {Latham},
  {Bakos}, {Bean}, {Berta-Thompson}, {Brown}, {Buchhave}, {Butler}, {Butler},
  {Chaplin}, {Charbonneau}, {Christensen-Dalsgaard}, {Clampin}, {Deming},
  {Doty}, {De Lee}, {Dressing}, {Dunham}, {Endl}, {Fressin}, {Ge}, {Henning},
  {Holman}, {Howard}, {Ida}, {Jenkins}, {Jernigan}, {Johnson}, {Kaltenegger},
  {Kawai}, {Kjeldsen}, {Laughlin}, {Levine}, {Lin}, {Lissauer}, {MacQueen},
  {Marcy}, {McCullough}, {Morton}, {Narita}, {Paegert}, {Palle}, {Pepe},
  {Pepper}, {Quirrenbach}, {Rinehart}, {Sasselov}, {Sato}, {Seager},
  {Sozzetti}, {Stassun}, {Sullivan}, {Szentgyorgyi}, {Torres}, {Udry}, \&
  {Villasenor}}]{ricker2014}
{Ricker}, G.~R., {Winn}, J.~N., {Vanderspek}, R., {et~al.} 2014, in Society of
  Photo-Optical Instrumentation Engineers (SPIE) Conference Series, Vol. 9143,
  Space Telescopes and Instrumentation 2014: Optical, Infrared, and Millimeter
  Wave, ed. J.~{Oschmann}, Jacobus~M., M.~{Clampin}, G.~G. {Fazio}, \& H.~A.
  {MacEwen}, 914320

\bibitem[{{Rieutord} {et~al.}(2005){Rieutord}, {Dintrans}, {Ligni{\`e}res},
  {Corbard}, \& {Pichon}}]{rieutord2005}
{Rieutord}, M., {Dintrans}, B., {Ligni{\`e}res}, F., {Corbard}, T., \&
  {Pichon}, B. 2005, in SF2A-2005: Semaine de l'Astrophysique Francaise, ed.
  F.~{Casoli}, T.~{Contini}, J.~M. {Hameury}, \& L.~{Pagani}, 759

\bibitem[{{Rieutord} \& {Espinosa Lara}(2013)}]{rieutord+espinosa2013}
{Rieutord}, M. \& {Espinosa Lara}, F. 2013, {Ab Initio Modelling of Steady
  Rotating Stars}, ed. M.~{Goupil}, K.~{Belkacem}, C.~{Neiner},
  F.~{Ligni{\`e}res}, \& J.~J. {Green}, Vol. 865, 49

\bibitem[{Rogers(2015)}]{rogers2015}
Rogers, T.~M. 2015, The Astrophysical Journal, 815, L30

\bibitem[{{Rogers} {et~al.}(2013){Rogers}, {Lin}, {McElwaine}, \&
  {Lau}}]{rogers2013}
{Rogers}, T.~M., {Lin}, D.~N.~C., {McElwaine}, J.~N., \& {Lau}, H.~H.~B. 2013,
  \apj, 772, 21

\bibitem[{{Saio}(1981)}]{saio1981}
{Saio}, H. 1981, \apj, 244, 299

\bibitem[{{Saio} {et~al.}(2018){Saio}, {Kurtz}, {Murphy}, {Antoci}, \&
  {Lee}}]{saio2018}
{Saio}, H., {Kurtz}, D.~W., {Murphy}, S.~J., {Antoci}, V.~L., \& {Lee}, U.
  2018, \mnras, 474, 2774

\bibitem[{{Spiegel} \& {Veronis}(1960)}]{Spiegel+Veronis1960}
{Spiegel}, E.~A. \& {Veronis}, G. 1960, \apj, 131, 442

\bibitem[{{Townsend}(2003)}]{townsend2003}
{Townsend}, R.~H.~D. 2003, \mnras, 340, 1020

\bibitem[{{Unno} {et~al.}(1989){Unno}, {Osaki}, {Ando}, {Saio}, \&
  {Shibahashi}}]{unno1989}
{Unno}, W., {Osaki}, Y., {Ando}, H., {Saio}, H., \& {Shibahashi}, H. 1989,
  {Nonradial oscillations of stars} (University of Tokyo Press)

\bibitem[{{Van Beeck} {et~al.}(2020){Van Beeck}, {Prat}, {Van Reeth}, {Mathis},
  {Bowman}, {Neiner}, \& {Aerts}}]{VanBeeck2020}
{Van Beeck}, J., {Prat}, V., {Van Reeth}, T., {et~al.} 2020, \aap, 638, A149

\bibitem[{{Van Reeth} {et~al.}(2018){Van Reeth}, {Mombarg, J. S. G.}, {Mathis,
  S.}, {Tkachenko, A.}, {Fuller, J.}, {Bowman, D. M.}, {Buysschaert, B.},
  {Johnston, C.}, {Garc\'{\i}a Hern\'andez, A.}, {Goldstein, J.}, {Townsend, R.
  H. D.}, \& {Aerts, C.}}]{vanreeth2018}
{Van Reeth}, T., {Mombarg, J. S. G.}, {Mathis, S.}, {et~al.} 2018, A\&A, 618,
  A24

\bibitem[{{Van Reeth} {et~al.}(2016){Van Reeth}, {Tkachenko}, \&
  {Aerts}}]{vanreeth2016}
{Van Reeth}, T., {Tkachenko}, A., \& {Aerts}, C. 2016, A\&A, 593, A120

\bibitem[{{Van Reeth} {et~al.}(2015){Van Reeth}, {Tkachenko}, {Aerts},
  {P{\'a}pics}, {Degroote}, {Debosscher}, {Zwintz}, {Bloemen}, {De Smedt},
  {Hrudkova}, {Raskin}, \& {Van Winckel}}]{vanreeth2015}
{Van Reeth}, T., {Tkachenko}, A., {Aerts}, C., {et~al.} 2015, \aap, 574, A17

\bibitem[{Wang {et~al.}(2016)Wang, Boyd, \& Akmaev}]{wang2016}
Wang, H., Boyd, J.~P., \& Akmaev, R.~A. 2016, Geoscientific Model Development,
  9, 1477

\end{thebibliography}

\appendix
\section{Derivation of the linearised momentum and continuity equations in the spheroidal basis}
\subsection{The momentum equation} \label{app:momentum_equation}
The linearised momentum equation for an inviscid uniformly rotating fluid is written as
\begin{equation}\label{eq:momentum_equation}
(\partial_t+\Omega\partial_\varphi) \vec{v}+ 2\vec{\Omega} \wedge \vec{v}=-\frac{1}{\rho_0}\vec{\nabla} P^{\prime}+\frac{\rho^{\prime}}{\rho_0 ^2}\vec{\nabla} P_0- \vec{\nabla} \Phi^{\prime}. \end{equation}
Using  the spheroidal basis $(\vec{a}_{\zeta}, \vec{a}_{\theta}, \vec{a} _ {\varphi})$ defined in Eq\;(\ref{eq:spheroidal_base}) we can express the velocity field as follows
\begin{align}
    \vec{v}&=v^{i} \vec{a}_{i}=v^{i} \alpha_i \vec{E}_{i}=v^{i} \alpha_i g_{ij}\vec{E}^{j},
\end{align}
with $\displaystyle{\alpha_\zeta= \frac{\zeta^{2}}{r^{2} r_{\zeta}}}$, $\displaystyle{\alpha_\theta=\frac{\zeta}{r^{2} r_{\zeta}}}$, $\displaystyle{\alpha_\varphi= \frac{\zeta}{r^{2} r_{\zeta} \sin \theta}}$, $(\vec{E}^{\zeta}, \vec {E}^{\theta}, \vec{E}^{\varphi})$ is the dual basis (the contravariant basis) and $g_{ij}$ is the covariant metric tensor defined as follows
\begin{equation}g_{i j}=\vec{E}_{i} \cdot \vec{E}_{j}=\left(\begin{array}{ccc}
r_{\zeta}^{2} & r_{\zeta} r_{\theta} & 0 \\
r_{\zeta} r_{\theta} & r^{2}+r_{\theta}^{2} & 0 \\
0 & 0 & r^{2} \sin ^{2} \theta
\end{array}\right).\end{equation}
We write also the uniform angular velocity in the natural covariant basis as
\begin{align}
  \vec{\Omega}&=\Omega(\cos{\theta}\vec{e_r}-\sin{\theta}\vec{e_\theta})\\
  &= \underbrace{\Omega\left(\frac{\cos{\theta}}{r_\zeta}+\frac{\sin{\theta}r_\theta}{r r_\zeta}\right)}_{=\Omega^\zeta}\vec{E}_\zeta+\underbrace{\left(-\frac{\Omega \sin{\theta}}{r}\right)}_{=\Omega^\theta}\vec{E}_\theta.
\end{align}
We then express the Coriolis acceleration in the spheroidal basis as follows
\begin{equation}
  2\vec{\Omega} \wedge \vec{v}= 2\varepsilon_{i l k} \Omega^{l} \alpha_k v^{k}  \vec{E}^i,
\end{equation}
where $\varepsilon^{i l k}$ is the Levi-Civita tensor
\begin{equation}
    \varepsilon_{i l k}=\sqrt{|g|}[i, l, k],
\end{equation}
with $|g|=\operatorname{det}\left(g_{i j}\right)=r^{4} r_{\zeta}^{2} \sin ^{2} \theta$ and
\begin{equation}
[i, l, k]=\left\{\begin{aligned}
+1 & \text { if }(i, l, k) \text { is an even permutation of }(\zeta,\theta,\varphi) \\
-1 & \text { if }(i, l, k) \text { is an odd permutation of }(\zeta,\theta,\varphi) \\
0 & \text { otherwise }
\end{aligned}.\right.
\end{equation}
Finally, we express the gradient of the scalar function $f=\{P^\prime,P_0,\Phi^\prime\}$  in the spheroidal basis as
\begin{equation}
    \vec{\nabla}f=\partial_i f \vec{E}^i.
\end{equation}
With these definitions, it is now possible to give an explicit expression of the momentum equation (Eq.\;\ref{eq:momentum_equation}) in the spheroidal basis
\begin{multline}
(\partial_t+\Omega\partial_\varphi) \alpha_j v^j g_{ji} + 2\varepsilon_{i l k} \Omega^{l} \alpha_k v^{k}  \\ =-\frac{1}{\rho_0}\partial_i P^{\prime} +\frac{\rho^{\prime}}{\rho_0 ^2}\partial_i P_0 - \partial_i \Phi^{\prime}; \end{multline}
for $i=\zeta$, $i=\theta$ and $i=\varphi$  we get the pseudo-radial (Eq.\;\ref{eq:radial_momentum}), the latitudinal (Eq.\;\ref{eq:latitudinal_momentum}), and the azimuthal (Eq.\;\ref{eq:azimuthal_momentum}) components of the momentum equation, respectively.

\subsection{The continuity equation} \label{app:continuity_equation}
The linearised continuity equation is given by 
\begin{equation}
\left(\partial_{t}+\Omega \partial_{\varphi}\right) \rho^\prime+\vec{\nabla} \cdot(\rho_0 \vec{v})=0,
\end{equation}
then by expressing the divergence in the spheroidal basis as
\begin{equation}
    \vec{\nabla} \cdot(\rho_0 \vec{v})=\frac{1}{\sqrt{|g|}} \partial_j\left(\sqrt{|g|} \rho_0 \alpha_j v^j\right), 
\end{equation}
we get the equation (\ref{eq:continuity}).

\section{Visual presentation of the terms involved in the derivation of the GLTE}\label{app:terms}
\begin{figure*}
    \centering
     \resizebox{\hsize}{!}{\includegraphics{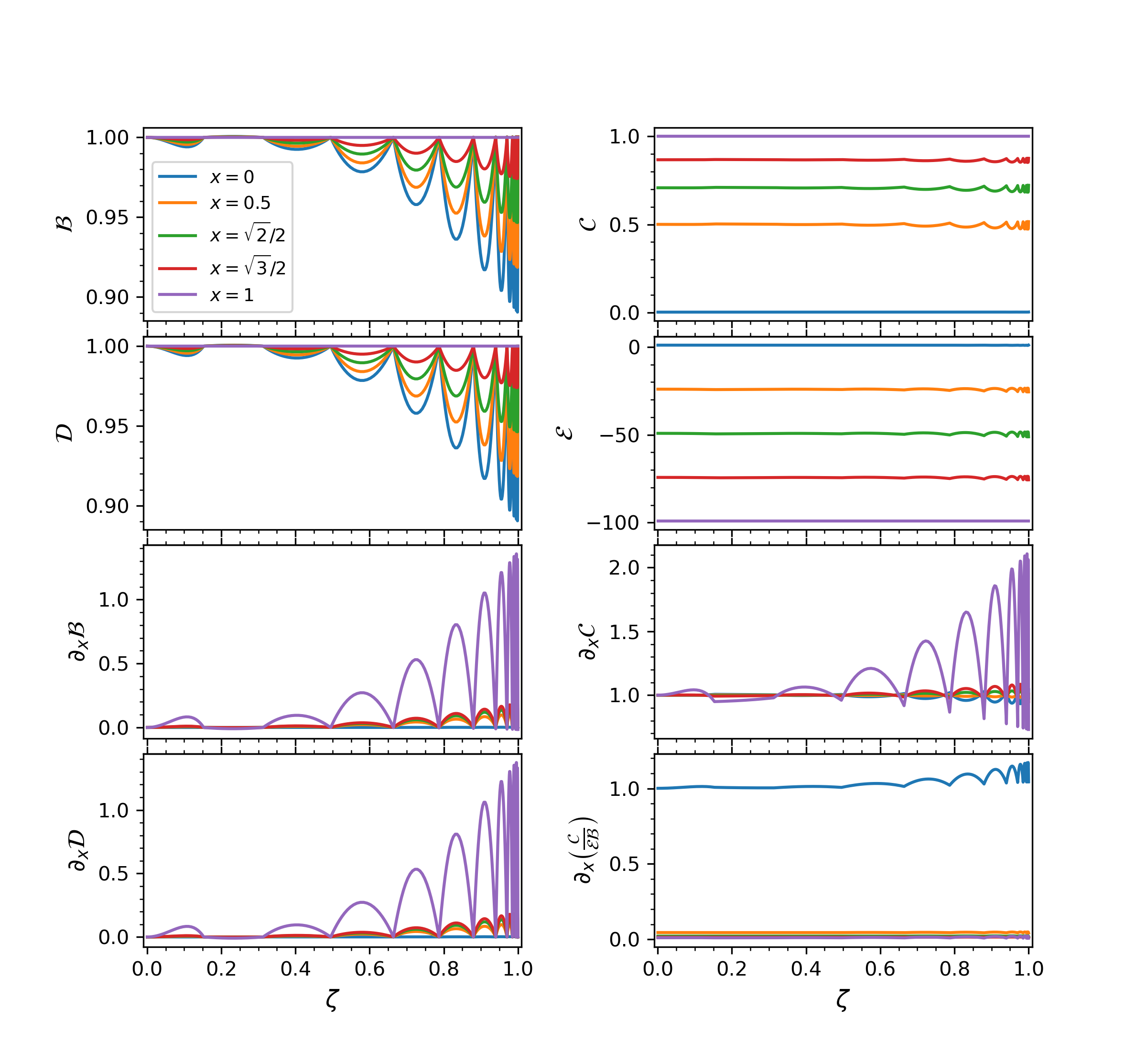}}
    \caption{Profiles of the coefficients $\mathcal{B}$, $\mathcal{C}$, $\mathcal{D}$, $\mathcal{E}$, $\partial_x\mathcal{B}$, $\partial_x\mathcal{C}$, $\partial_x\mathcal{D}$ and $\partial_x\left(\displaystyle{\frac{\mathcal{C}}{\mathcal{E}\mathcal{B}}}\right)$ involved in the derivation of the GLTE (Eq.\;\ref{eq:glte}) as a function of $\zeta$ at different reduced colatitudes $x$ using an ESTER model $(\mathrm{M}=3 \mathrm{M}_{\odot},~X_{\rm c} = 0.7)$ rotating at $20\%$ of the Keplerian break-up rotation rate for a spin parameter $\nu=10$.}
    \label{fig:terms_glte}
\end{figure*}

We represent in Fig\;\ref{fig:terms_glte} the profiles of the 8 terms ($\mathcal{A}$ is already represented in the top panel of Fig\;\ref{fig:A-0.6}) involved in the derivation of the GLTE (Eq.\;\ref{eq:glte}) in the general case of spheroidal geometry presented in the Table\;\ref{table:termes} as a function of the pseudo-radius $\zeta$ at different reduced colatitudes $x$. In the main text (Sect.\;\ref{subsect:validity}), we mainly focused on the coefficient $\mathcal{A}$ and the Brunt-Vaïsälä frequency $N$ because they are the important quantities to be able to build the generalised TAR formalism. The oscillating behaviour of some terms is caused by the fact that the mapping (Eq.\;\ref{eq:mapping}) is different in each subdomain of the model which is continuous and derivable, but not twice differentiable with respect to $\zeta$ between subdomains.

\end{document}